\documentclass[final,1p,times,12pt]{elsarticle}

\journal{Computer Methods in Applied Mechanics and Engineering}

\usepackage{amsfonts,setspace}
\usepackage{amsthm}
\usepackage{amsmath}
\usepackage{amssymb}
\usepackage{color}
\usepackage{graphicx}
\usepackage{float}
\usepackage{mathrsfs}
\usepackage{mathtools}
\usepackage{latexsym}
\usepackage{subfig}
\usepackage{bm} 

\usepackage{geometry}
\usepackage{booktabs}

\usepackage{algorithm}
\usepackage[noend]{algpseudocode}

\usepackage{xfrac} 
\usepackage{flexisym}
\usepackage{lineno,hyperref}
\modulolinenumbers[5]
\hypersetup{
    colorlinks,
    citecolor=blue,
    filecolor=magenta,
    linkcolor=red,
}

\newtheorem{Theorem}{Theorem}

\def\HH{\mathcal{H}}
\def\LL{\mathcal{L}}
\def\KK{\mathcal{K}}

\def\RR{\mathcal{R}}

\def\II{\mathcal{I}}
\def\PP{\textbf{P}}
\def\AA{\textbf{A}}

\def\SS{\textbf{S}}
\def\Id{\textbf{I}}

\def\xx{\text{\bf x}}
\def\cc{\text{\bf c}}
\def\yy{\text{\bf y}}
\def\nn{\text{\bf n}}
\def\zz{\text{\bf z}}

\begin{document}

\begin{frontmatter}




\title{\textbf{Local on-surface radiation condition for multiple scattering of waves from convex obstacles}}
\author[label,label2]{Sebasti\'{a}n Acosta\corref{cor1}}
\ead{sebastian.acosta@bcm.edu}
\ead[url]{https://sites.google.com/site/acostasebastian01}
\cortext[cor1]{Corresponding author}
\address[label]{Department of Pediatrics, Baylor College of Medicine, Houston TX, USA}
\address[label2]{Predictive Analytics Laboratory, Texas Children's Hospital, Houston TX, USA}

\begin{abstract}
We propose a novel on-surface radiation condition to approximate the outgoing solution to the Helmholtz equation in the exterior of several impenetrable convex obstacles. Based on a local approximation of the Dirichlet--to--Neumann operator and a local formula for wave propagation, this new formulation simultaneously accounts for the outgoing behavior of the solution as well as the reflections arising from the multiple obstacles. The method involves tangential derivatives only, avoiding the use of integration over the surfaces of the obstacles. As a consequence, the method leads to sparse matrices and $\mathcal{O}(N)$ complexity. Numerical results are presented to illustrate the performance and limitations of the proposed formulation. Possible improvements and extensions are also discussed.
\end{abstract}

\begin{keyword}
On-surface radiation conditions \sep semi-analytical approximations \sep absorbing boundary conditions \sep wave scattering \sep Helmholtz equation
\MSC[2010] 35J05 \sep 58J40 \sep 58J05 \sep 41A60 \sep 65N38
\end{keyword}


\end{frontmatter}


\section{Introduction} \label{Section.Intro}

The on-surface radiation conditions (OSRCs), originally developed by Kriegsmann, Taflove and Umashankar \cite{Kriegsmann1987}, are semi-analytical formulations to roughly approximate the solution of wave scattering problems. The approximate nature of an OSRC prevents it from being used as a satisfactory solution procedure. However, this semi-analytical construct yields tremendous computational speed which is useful to explore feasibility regions for optimization problems, to produce a good initial guess for iterative methods, and to design inexpensive preconditioners to solve boundary integral equations (BIE) numerically using fixed-point or Krylov subspace methods \cite{Antoine2004,Antoine2005,Antoine2007,Darbas2013}. The OSRCs have been studied by many researchers including Antoine and collaborators who formulated them in a differential geometric setting in order to handle non-canonical surfaces \cite{Antoine1999,Antoine2001,
Antoine2001b,Antoine2006,Antoine2008}. Other improvements have been accomplished by Jones \cite{Jones1988,Jones1990,Jones1992}, Ammari \cite{Ammari1998,Ammari1998b}, Calvo et al. \cite{Calvo2004,Calvo2003}, Barucq et al. \cite{Barucq2003,Barucq2010a,Barucq2012}, Chaillat et al. \cite{Chaillat2015,Chaillat2017} and Darbas et al. \cite{Antoine2004,Antoine2005,Antoine2007,Darbas2013,Darbas2006}. See also  \cite{Atle2007,Acosta2015f,Stupfel1994,Murch1993,Teymur1996,
Medvinsky2010,Chniti2016,Alzubaidi2016,Acosta2015f,Acosta2017c}.

The underlying assumption in most of the aforementioned works is that the solution radiates outwardly at every point on the surface of the scatterer. This assumption may be violated when the scatterer is not convex since the wave field may exhibit reflections from the scatterer to itself. Hence, one of the main drawbacks of the OSRC method is that its formulation and accuracy are limited to convex scatterers. This issue was addressed by the author in \cite{Acosta2015f} and by Alzubaidi, Antoine and Chniti in \cite{Alzubaidi2016} for multiple scattering problems where the scatterer consists of several disjoint obstacles. The formulations in \cite{Acosta2015f,Alzubaidi2016} successfully account for the propagation of waves from one obstacle to another. However, the wave propagation formulas are non-local. Due to the use of boundary integral operators, integration of the wave fields over the surface of the obstacles is required. As a consequence, matrices with dense blocks are obtained upon discretization, which leads to high memory demands and costly matrix inversion, especially in the three-dimensional setting and also for high frequencies.

Our main goal is to modify the formulation of the OSRC for multiple obstacles in such a way that its discretization leads to sparse matrices. This is accomplished by defining a local wave propagation formula to transmit the influence of one obstacle on another using differential rather than integral operators. In Section \ref{Section.MultipleScatt} we provided the mathematical formulation of the multiple scattering problem and decompose it into a system of single-scattering problems. Based on \cite{Antoine1999,Acosta2015f,Acosta2017c}, we derive a local formula for the propagation of waves which allows us to account for the wave reflections between obstacles. This formula is developed in Section \ref{Section.Propagation}. In Section \ref{Section.FFP} we use the propagation formula to evaluate the far-field pattern using local operations as well.
In Section \ref{Section.Numerics}, we propose how to numerically implement the OSRC based on a triangulation of the surfaces. A few numerical results and comparisons with exact solutions are shown as well. In Section \ref{Section.Precond}, we implement the OSRC as a preconditioner to reduce the spectral radius of a block-Jacobi iteration matrix to solve a multiple scattering problem using a boundary element method. 
Limitations, future work and conclusions are discussed in Section \ref{Section.Conlusion}.

\section{Mathematical formulation} \label{Section.MultipleScatt}

We seek to approximate an outgoing solution to the Helmholtz equation in the exterior to a set of $J$ impenetrable convex obstacles. Each obstacle is enclosed by a closed smooth surface $\partial \Omega_{j}$ that separates the space into a simply connected bounded domain $\Omega_{j}^{-}$ and an exterior domain $\Omega_{j}^{+}$. 
We also define
\begin{align} \label{Eqn.DomainDef}
\Omega^{-} = \bigcup_{j=1}^{J} \Omega_{j}^{-}, \qquad \Omega^{+} = \bigcap_{j=1}^{J} \Omega_{j}^{+} \qquad \text{and} \qquad \partial \Omega = \bigcup_{j=1}^{J} \partial \Omega_{j}.
\end{align}

The wave field is assumed to satisfy the following boundary value problem,
\begin{subequations} \label{Eqn.Main}
\begin{align} 
& \Delta u + k^2 u = 0  \qquad  \text{in $\Omega^{+}$},  \label{Eqn.Main1} \\
& u = f   \qquad \qquad \text{on $\partial \Omega$},  \label{Eqn.Main2} \\
& \lim_{r \to \infty} r \left( \partial_{r}u - \iota k u \right) = 0,  \label{Eqn.Main3}
\end{align}
\end{subequations}
where the wavenumber $k > 0$ is constant, $\iota$ is the imaginary unit, $r = |\xx|$, and $\partial \Omega$ is the boundary of $\Omega^{+}$. The proposed approach to roughly approximate the solution $u$ of the boundary value problem (\ref{Eqn.Main}) is based on the following theorem whose objective is to express the solution $u$ as the sum of purely-outgoing wave fields $u_{j}$ each radiating from a single surface $\partial \Omega_{j}$ for $j=1,2,...,J$. A proof is found in \cite{Balabane2004}.

\begin{Theorem} \label{Thm.Decomp}
Let the closures of $\Omega_{j}^{-}$ and of $\Omega_{i}^{-}$ be mutually disjoint for $i \neq j$, and let $u$ be a radiating solution to the Helmholtz equation in $\Omega^{+}$. Then $u$ can be uniquely decomposed into purely-outgoing wave fields $u_{j}$ for $j=1,2,...,J$ such that
\begin{align} \label{Eqn.Decomp}
u = \sum_{j=1}^{J} u_{j}, \qquad \text{in $\Omega^{+}$},
\end{align}
where $u_{j}$ radiates only from $\partial \Omega_{j}$, that is,
\begin{align} \label{Eqn.PureRad}
\Delta u_{j} + k^2 u_{j} = 0 \text{ in $\Omega_{j}^{+}$,} \qquad \text{and} \qquad \lim_{r \to \infty} r \left( \partial_{r} u_{j} - \iota k u_{j} \right) = 0.
\end{align}
\end{Theorem}

Using Green's theorem for each of the purely-outgoing wave fields $u_{j}$, we obtain a representation of the solution
\begin{align} \label{Eqn.Green}
u(\xx) = \sum_{j=1}^{J} \int_{\partial \Omega_{j}} \left[ \partial_{\nu(\yy)} \Phi(\xx,\yy) u_{j}(\yy) - \Phi(\xx,\yy) \partial_{\nu} u_{j}(\yy) \right] dS(\yy), \qquad \xx \in \Omega^{+},
\end{align}
where $\Phi$ is the fundamental solution to the Helmholtz equation. If the exact outgoing Dirichlet-to-Neumann (DtN) map $\Lambda_{j}$ on each surface $\partial \Omega_{j}$ is available, then $\partial_{\nu}u_{j} = \Lambda_{j} u_{j}$ and we obtain
\begin{align} \label{Eqn.GreenDtN}
u(\xx) = \sum_{j=1}^{J} \int_{\partial \Omega_{j}} \left[ \partial_{\nu(\yy)} \Phi(\xx,\yy) - \Phi(\xx,\yy) \Lambda_{j} \right] u_{j}(\yy) dS(\yy), \qquad \xx \in \Omega^{+}.
\end{align}
The representation (\ref{Eqn.GreenDtN}) renders the solution in $\Omega^{+}$ provided that the Dirichlet data of each purely-outgoing wave field $u_{j}$ is known. This data can be found by imposing the boundary condition (\ref{Eqn.Main2}) on each boundary $\partial \Omega_{j}$ which leads to the following system of equations,
\begin{align} \label{Eqn.MainSystem}
u_{i} + \sum_{j \neq i} \PP_{i,j} u_{j} = f_{i} \quad \text{on $\partial \Omega_{i}$}, \qquad i=1,2,...,J
\end{align}
where $f_{i}$ is the evaluation of $f$ on the surface $\partial \Omega_{i}$, and where the propagation of the wave field $u_{j}$ from the surface $\partial \Omega_{j}$ to the surface $\partial \Omega_{i}$ for $j\neq i$ is given by the following operator,
\begin{align} \label{Eqn.Propagator}
\left(\PP_{i,j} u_{j}\right)(\xx) = \int_{\partial \Omega_{j}} \left[ \partial_{\nu(\yy)} \Phi(\xx,\yy) -  \Phi(\xx,\yy) \Lambda_{j} \right] u_{j}(\yy) dS(\yy), \qquad \xx \in \partial \Omega_{i}.
\end{align}
For practical purposes, the evaluation of the propagation operator in (\ref{Eqn.Propagator}) has two main challenges. First, the DtN map $\Lambda_{j}$ must be evaluated or at least roughly approximated which is the objective of OSRC in general. Second, the propagation operator involves integration over the boundary  of each obstacle. This leads to the appearance of dense blocks once the governing system (\ref{Eqn.MainSystem}) is discretized. Our objective in the following section is to mitigate this latter problem.

\section{Local formula for propagation of wave fields} \label{Section.Propagation}

We seek to derive an analytical formula to roughly approximate the propagation of a wave field away from a generic convex surface $\Gamma$. It is conceptually similar to the beam propagation method developed in \cite{Antoine2010} for two-dimensional scattering problems.
This formula will allow us to account for the interaction between obstacles in order to approximate the system (\ref{Eqn.MainSystem}) using sparse matrices. This formula is for the propagation of waves on a expansive foliation of parallel surfaces generated by the smooth surface $\Gamma$. We follow closely the approach from \cite{Antoine1999}. The domain exterior to $\Gamma$ is denoted $\Omega$. This family of surfaces $\Gamma_{s}$ parametrized by $s \geq 0$ is defined as follows
\begin{align} \label{Eqn.Parallel_Surfaces}
\Gamma_{s} = \{ \yy = \xx + s \nn(\xx) \,:\, \xx \in \Gamma \}  
\end{align}
where $\nn(\xx)$ is the outward normal vector of the surface $\Gamma$ at the point $\xx \in \Gamma$. Notice $\Gamma = \Gamma_{0}$. Since $\Gamma$ is convex, the family of surfaces $\Gamma_{s}$ foliates $\Omega$. For any $\yy \in \Omega$ there exists unique $\xx \in \Gamma$ and unique $s > 0$ such that $\yy = \xx + s \nn(\xx)$. Moreover, the outward normal vector $\nn(\yy)$ of the surface $\Gamma_{s}$ at a point $\yy = \xx + s \nn(\xx)$ coincides with the normal vector $\nn(\xx)$. See details in \cite[\S 6.2]{Kress-Book-1999}, \cite[\S 14.6]{Gil-Tru-2001}, \cite[Probl. 11 \S 3.5]{DoCarmo1976} and \cite{Antoine1999}. The point $\xx = \xx(\yy)$ can be regarded as the projection of $\yy$ on the surface $\Gamma$, and $s=s(\yy)$ is the distance from $\xx$ to $\yy$ \cite[Ch. 2]{Deu-2001}. The pair $(\xx,s)$ satisfies
\begin{align} \label{Eqn.Minimizer}
\xx = \operatorname*{argmin}_{\zz \in \Gamma} \, | \zz - \yy | \quad \text{and} \quad s = \operatorname*{min}_{\zz \in \Gamma} \, | \zz - \yy |.
\end{align}

Now we proceed to write the Helmholtz differential operator in terms of a tangential system of coordinates on the surface $\Gamma_{s}$
\begin{align} \label{Eqn.LaplacianLocal}
\LL(s) w = \Delta w + k^2 w = \partial_{s}^2 w + 2 \HH_{s}  \partial_{s} w + \Delta_{\Gamma_{s}} w + k^2 w.
\end{align}
Here $\HH_{s}$ is the mean curvature of the surface $\Gamma_{s}$ and $\Delta_{\Gamma_{s}}$ is the Laplace--Beltrami operator on the same surface. The differential $\partial_{s}$ represents the derivative in the outward normal direction on $\Gamma_{s}$. See \cite{Antoine1999} for a concise review of the differential geometry of surfaces, including the definition of curvatures and the Laplace--Beltrami operator. The mean curvature $\HH_{s}$ and Gauss curvature $\KK_{s}$ of the surface $\Gamma_{s}$ satisfy the following relations,
\begin{align} \label{Eqn.Curvatures}
& \HH_{s}(\xx) = \frac{ \HH(\xx) + s \KK(\xx) }{\mu_{s}(\xx)} \quad \text{and} \quad \KK_{s}(\xx) = \frac{\KK(\xx)}{\mu_{s}(\xx)} 
\end{align}
where the symbol $\mu_{s}$ stands for
\begin{align} \label{Eqn.MU}
\mu_{s}(\xx) = \det \left(\II + s \RR(\xx) \right) = 1 + 2 s \HH(\xx) + s^2 \KK(\xx). 
\end{align}
Here $\HH$ and $\KK$ are mean and Gauss curvature of the surface $\Gamma$, respectively. Also, $\RR$ is the curvature tensor and $\II$ is the identity on the tangent plane. The curvature tensor $\RR$ can be represented as a self-adjoint matrix with eigenvalues $\kappa_{1}$ and $\kappa_{2}$, called the principal curvatures, such that
\begin{align*} 
\HH = \frac{\kappa_{1} + \kappa_{2}}{2} \quad \text{and} \quad \KK = \kappa_{1} \kappa_{2}.
\end{align*}
The Laplace-Beltrami operator satisfies
\begin{align} \label{Eqn.LB}
\Delta_{\Gamma_{s}}w = \mu_{s}^{-1}  \text{div}_{\Gamma} \left( \mu_{s} \left( \II + s \RR  \right)^{-2} \text{grad}_{\Gamma} w \right) \approx \mu_{s}^{-1} \Delta_{\Gamma} w
\end{align}
where this latter approximation is valid when $\kappa_{1} \approx \kappa_{2}$ so that $\mu_{s} \left( \II + s \RR  \right)^{-2} \approx \II$.

Now we seek to decompose the wave field propagating across a surface $\Gamma_{s}$ into incoming and outgoing components using Nirenberg's factorization theorem \cite{Nirenberg1973,Antoine1999}. There are two pseudo--differential operators $\Lambda^{-}(s)$ and $\Lambda^{+}(s)$ of order $+1$, such that  
\begin{align} \label{Eqn.Decomp01}
\LL(s) w = (\partial_{s} - \Lambda^{-}(s))(\partial_{s} - \Lambda^{+}(s)) w.
\end{align}
We refer to $\Lambda^{+}(s)$ and $\Lambda^{-}(s)$ as the outgoing and incoming DtN operators for the radiating boundary value problem defined in the exterior of $\Gamma_{s}$. The operator $\Lambda_{s}^{+}$ admits the following  second order approximation derived in \cite{Antoine1999}, 
\begin{align} 
\Lambda^{+}(s) = \iota k - \HH_{s} - \frac{1}{2 \iota k} \left( \Delta_{\Gamma_{s}} + \HH_{s}^2 - \KK_{s}  \right) + \mathcal{O}(k^{-2}). \label{Eqn.Lp1} 
\end{align}

If $w$ is purely-outgoing from the surface $\Gamma$, then $\partial_{s} w = \Lambda^{+}(s) w$. Using the approximation (\ref{Eqn.Lp1}) for the outgoing DtN operator, we obtain that $w$ satisfies the following differential equation in $s > 0$,
\begin{align}  \label{Eqn.DiffEqn}
\frac{\partial w}{\partial s} \approx \iota k w - \HH_{s} w - \frac{1}{2 \iota k} \left( \Delta_{\Gamma_{s}} + \HH_{s}^2 - \KK_{s} \right) w.
\end{align}
This is a separable differential equation. In order to solve it approximately in closed-form, we need the following expressions
\begin{align*} 
& \int_{0}^{s} \HH_{z} dz = \frac{1}{2} \ln \mu_{s} \\
& \int_{0}^{s} \left[ \HH_{z}^2 - \KK_{z} \right] dz \approx - \frac{1}{2} \left[ \frac{s \KK}{1 + s \HH} + \HH_{s} - \HH  \right]  \\
& \int_{0}^{s} \Delta_{\Gamma_{z}} dz \approx \frac{ s }{ 1 + s \HH } \Delta_{\Gamma}
\end{align*}
where we have employed the approximation in (\ref{Eqn.LB}) for the third integral. Hence, we obtain
\begin{align*} 
\ln \left( \frac{w}{w_{\rm o}}\right) & \approx \iota k s - \frac{1}{2} \ln \mu_{s} - \frac{1}{2 \iota k} \left( \frac{s }{1+s\HH} \Delta_{\Gamma} - \frac{1}{2} \left[ \frac{s \KK}{1 + s \HH} + \HH_{s} - \HH  \right]    \right)
\end{align*}
which leads to
\begin{align} 
w(s) \approx \frac{e^{\iota k s}}{\mu_{s}^{1/2} } \exp \left( \frac{1}{4 \iota k} \left[ \frac{s \KK}{1 + s \HH} + \HH_{s} - \HH  \right] \right)   \exp  \left(  \frac{- s \, \Delta_{\Gamma} }{ 2 \iota k \, \left( 1 + s \HH \right) }   \right) w_{\rm o} \label{Eqn.WaveField}
\end{align}
where we have neglected the tangential derivatives of $\HH$ and $\KK$ in order to factor the exponential. Here $w_{\rm o}$ represents the trace of $w$ on the boundary $\Gamma$. In order to maintain the locality of the evaluation, we further approximate the exponential of the Laplace--Beltrami operator using an implicit linear approximation valid for large frequency $k$,
\begin{align} 
 \exp  \left(  \frac{ - s \, \Delta_{\Gamma} }{ 2 \iota k \, \left( 1 + s \HH \right) }   \right) \approx \left( 1 + \frac{ s \, \Delta_{\Gamma} }{ 2 \iota k \, \left( 1 + s \HH \right) } \right)^{-1}. \label{Eqn.ApproxExp}
\end{align}
This implicit approximation is chosen to preserve the stability (boundedness) of the operations. In fact, the spectrum of the operator in (\ref{Eqn.ApproxExp}) lies in the unit circle because the Laplace-Beltrami $\Delta_{\Gamma}$ operator has real eigenvalues.

Now we go back to the multi-scattering problem with its notation formulated in Section \ref{Section.MultipleScatt}. In order to propagate the wave field $u_{j}$ from the surface $\partial \Omega_{j}$ onto the surface $\partial \Omega_{i}$, we approximate the propagator (\ref{Eqn.Propagator}) as follows, 
\begin{align} \label{Eqn.PropagatorD}
(\PP_{i,j} u_{j} )(\yy) \approx \frac{e^{\iota k s}}{\mu_{s}(\xx)^{1/2} } \exp \left( \frac{1}{4 \iota k} \left[ \frac{s \KK(\xx)}{1 + s \HH(\xx)} + \HH_{s}(\xx) - \HH(\xx) \right] \right)   \left( 1 + \frac{s \, \Delta_{\partial \Omega_{j}} }{ 2 \iota k \, \left( 1 + s \HH(\xx) \right) }  \right)^{-1} u_{j}(\xx)
\end{align}
where $\yy =\xx + s \nn(\xx)$. Due to the assumed convexity of $\partial \Omega_{j}$, the point $\xx \in \partial \Omega_{j}$ and $s>0$ are determined uniquely by $\yy \in \partial \Omega_{i}$ satisfying (\ref{Eqn.Minimizer}). The curvatures $\HH$ and $\KK$, and the Laplace--Beltrami operator $\Delta_{\partial \Omega_{j}}$ correspond to the surface $\partial \Omega_{j}$. Regarding the boundedness of (\ref{Eqn.PropagatorD}), notice that the two exponentials have purely imaginary arguments, the operator (\ref{Eqn.ApproxExp}) is non-expansive, and the factor $\mu_{s}(\xx) > 1$ for $s>0$ and for convex $\partial \Omega_{j}$ ($\HH > 0$, $\KK >0$). Therefore each propagator $\PP_{ij}$ can be shown to be a contraction mapping. This renders the OSRC system (\ref{Eqn.MainSystem}) uniquely solvable by the Neumann series theorem in an $L^2$ framework.

\section{Far-field pattern} \label{Section.FFP}

It is commonly of interest to evaluate the radiating field $u$ away from the obstacles. The asymptotic behavior  is characterized by the far-field pattern $u^{\infty}$ of the radiating field $u$ which is given by
\begin{align} \label{Eqn.FFasympt}
u(\yy) = \frac{e^{i k r}}{r} u^{\infty}\left(\hat{\yy}\right) + \mathcal{O}\left( r^{-1} \right), \qquad \text{as $r \to \infty$,}
\end{align}
where $r = |\yy|$ and $\hat{\yy} = \yy / |\yy|$. From the superposition of the purely-outgoing fields, we find that
\begin{align} \label{Eqn.FFPdef}
u^{\infty} = \sum_{j=1}^{J} u_{j}^{\infty}
\end{align}
so that it only remains to find the far-field pattern $u_{j}^{\infty}$ of each purely-outgoing field $u_{j}$. We proceed with the following asymptotic expansions valid as $r \to \infty$,
\begin{subequations} \label{Eqn.Asymptotics}
\begin{align} 
s^2 &= r^2 - 2 r \xx \cdot \hat{\yy} + |\xx|^2 \\
s &= r - \xx \cdot \hat{\yy} + \mathcal{O}\left( r^{-1} \right) \\
\frac{1}{\mu_{s}^{1/2}} &= \frac{1}{r \KK^{1/2}}  + \mathcal{O}\left( r^{-2} \right) \\
\frac{s \KK}{1 + s \HH} + \HH_{s} - \HH &= \frac{\KK - \HH^2}{\HH} + \mathcal{O}\left( r^{-1}\right) \\
\frac{ s }{1 + s \HH } &= \frac{1}{\HH} + \mathcal{O}\left( r^{-1} \right)
\end{align}
\end{subequations} 
Using (\ref{Eqn.Asymptotics}) and (\ref{Eqn.WaveField}), we find that the purely-outgoing wave field $u_{j}$ has the following asymptotic behavior
\begin{align*} 
u_{j}(r \hat{\yy}) = \frac{e^{\iota k r}}{r} \left[ \frac{e^{- \iota k \xx \cdot \hat{\yy}}}{\KK(\xx)^{1/2}} \exp \left( \frac{\KK(\xx) - \HH(\xx)^2}{4 \iota k \, \HH(\xx)}   \right)   \left(1 +  \frac{\Delta_{\partial \Omega_{j}} }{ 2 \iota k \, \HH(\xx) }   \right)^{-1} u_{j}(\xx) + \mathcal{O}\left( r^{-1} \right)  \right]
\end{align*}
which implies that the far-field pattern corresponding to $u_{j}$ is given by
\begin{align} \label{Eqn.FFP_uj}
u^{\infty}_{j}(\hat{\yy}) = \frac{e^{- \iota k \xx \cdot \hat{\yy}}}{\KK(\xx)^{1/2}} \exp \left( \frac{\KK(\xx) - \HH(\xx)^2}{4 \iota k \, \HH(\xx)}   \right)   \left(1 + \frac{\Delta_{\partial \Omega_{j}} }{ 2 \iota k \, \HH(\xx) } \right)^{-1} u_{j}(\xx) 
\end{align}
where $\xx \in \partial \Omega_{j}$ is uniquely determined by $\hat{\yy}$ such that $\nn(\xx) = \hat{\yy}$. Hence, once the Dirichlet data of the field $u_{j}$ is found by solving the system (\ref{Eqn.MainSystem}), then plugging (\ref{Eqn.FFP_uj}) into (\ref{Eqn.FFPdef}) renders an approximation for the far-field pattern $u^{\infty}$.

\section{Discrete implementation on triangulated surfaces} \label{Section.Numerics}

In this section, we propose a numerical implementation of the on-surface radiation condition defined by the system (\ref{Eqn.MainSystem}) where the propagation operator $\PP_{i,j}$ is roughly approximated by (\ref{Eqn.PropagatorD}). The approach is based on discretizing the Laplace--Beltrami operator, the mean and Gauss curvatures using a triangulation of the surfaces $\partial \Omega_{j}$ for $j=1,2,...,J$. Our guiding references for approximating geometrical properties and operators on triangulated surfaces are \cite{Xu2006a,Xu2004,Xu2004a,Meyer2003,Magid2007,Bobenko2007}.

\subsection{Discrete Laplace--Beltrami operator and curvatures} \label{Subsection.LBCurv}

The discrete Laplace--Beltrami operator is described as follows. Let $\{\xx_{j}\}_{j}^{J}$ be the collection of vertices of the triangulation $\mathcal{T}$ of the surface $\Gamma$. For a fixed vertex $\xx_{j}$, let $\{\xx_{j(i)}\}_{i=1,2,...}$ be the neighboring vertices of $\xx_{j}$, and let
$\{e_{j(i)}\}_{i=1,2,...}$ be the edges of the triangulation that connect $\xx_{j}$ and $\xx_{j(i)}$. 
Now for each edge $e_{j(i)}$, let $\alpha_{j(i)}$ and $\beta_{j(i)}$ be the angles opposing the edge $e_{j(i)}$ in the two triangles that share that edge. For a smooth function $u$ defined on the surface $\Gamma$, we use the following discrete Laplace--Beltrami operator,
\begin{align} \label{Eqn.DiscreteLB}
(\Delta_{\Gamma} u)(\xx_{j}) = \frac{1}{2} \sum_{i} \frac{ \cot \alpha_{j(i)}  + \cot \beta_{j(i)}}{ \left( A_{j} A_{j(i)} \right)^{1/2} } \left( u(\xx_{j(i)}) - u(\xx_{j}) \right),
\end{align}
where $A_{j}$ is the area associated with the vertex $\xx_{j}$ defined as \sfrac{1}{3} of the total area of the triangular elements sharing the vertex $\xx_{j}$. 

For the discrete mean curvature, we first define the discrete mean curvature vector as
\begin{align} \label{Eqn.DiscreteMCV}
\textbf{H} (\xx_{j}) = - \frac{1}{4} \sum_{i} \frac{ \cot \alpha_{j(i)}  + \cot \beta_{j(i)}}{ \left( A_{j} A_{j(i)} \right)^{1/2} } \left( \xx_{j(i)} - \xx_{j} \right).
\end{align}
Then the mean curvature is given by
\begin{align} \label{Eqn.DiscreteMC}
\HH (\xx_{j}) = \textbf{H} (\xx_{j}) \cdot \textbf{n}(\xx_{j})
\end{align}
where $\textbf{n}(\xx_{j})$ is the normal vector at the vertex $\xx_{j}$.

The discrete Gauss curvature at each vertex $\xx_{j}$ of the triangulation, is defined as follows  \cite{Xu2006a,Magid2007,Surazhsky2003}. Let $\theta_{i}$ be the angle between two successive edges sharing the vertex $\xx_{j}$. Then, the discrete Gauss curvature is given by
\begin{align} \label{Eqn.DiscreteGaussCurv}
\KK(\xx_{j}) = \frac{2 \pi - \sum_{i} \theta_{i}}{A_{j}}
\end{align}
where $A_{j}$ is the area associated with the vertex $\xx_{j}$ as defined above.

\subsection{Discrete wave propagator} \label{Subsection.DiscretePropagator}

Given the triangulation $\mathcal{T}_{j}$ of the surface $\partial \Omega_{j}$, we have the necessary definitions in Subsection \ref{Subsection.LBCurv} to implement a discrete version of the propagator (\ref{Eqn.PropagatorD}) to pose the system (\ref{Eqn.MainSystem}) at the discrete level. The discrete operator $\PP_{i,j}$ propagates the waves from the triangulation $\mathcal{T}_{j}$ of the surface $\partial \Omega_{j}$ to the triangulation $\mathcal{T}_{i}$ of the surface $\partial \Omega_{i}$. Let the triangulations $\mathcal{T}_{j}$ and $\mathcal{T}_{i}$ have $N_{j}$ and $N_{i}$ vertices, respectively. Then the propagation operator can be represented as a matrix $\PP_{i,j} : \mathbb{R}^{N_{j}} \to \mathbb{R}^{N_{i}}$. We proceed to describe the construction of this matrix. Let $\yy_{n} \in \mathcal{T}_{i}$ where $n$ is the index in the list of vertices of $\mathcal{T}_{i}$. Then we can find $\xx_{m} \in \mathcal{T}_{j}$ and $s_{m} > 0$ as the discrete version of (\ref{Eqn.Minimizer}), that is,
\begin{align} \label{Eqn.MinimizerDiscrete}
m = \operatorname*{argmin}_{1 \leq p \leq M_{j}} \, | \xx_{p} - \yy_{n} | \quad \text{and} \quad s_{m} = | \xx_{m} - \yy_{n} |.
\end{align}
Then the matrix $\PP_{i,j}$ can be written as
\begin{align} \label{Eqn.MatrixForm}
\PP_{i,j} = \AA_{i,j} \left( \Id + \frac{\Delta_{\mathcal{T}_{j}}}{2 \iota k  \HH }  \right)^{-1}
\end{align}
where $\Delta_{\mathcal{T}_{j}}$ is an $N_{j} \times N_{j}$ matrix defined by (\ref{Eqn.DiscreteLB}) which approximates the Laplace--Beltrami operator, and $\AA_{i,j}$ is an $N_{i} \times N_{j}$ matrix. Each row of this matrix has exactly one non-zero entry. For the $n^{\text{th}}$ row, this non-zero entry is at the $m^{\text{th}}$ column and it is given by
\begin{align} \label{Eqn.Entries}
(\AA_{i,j})_{n,m} &= \frac{e^{\iota k s_{m}}}{\mu_{s_{m}}(\xx_{m})^{1/2} } \exp \left( \frac{1}{4 \iota k} \left[ \frac{s_{m} \KK(\xx_{m})}{1 + s_{m} \HH(\xx_{m})} + \HH_{s_{m}}(\xx_{m}) - \HH(\xx_{m}) \right] \right)
\end{align}
where $m$ depends on $n$ by satisfying (\ref{Eqn.MinimizerDiscrete}). With this notation, we can write the discrete version of the system (\ref{Eqn.MainSystem}) in block-matrix form as follows,
\begin{align} \label{Eqn.SystemMatrix}
\begin{bmatrix}
    \Id_{11} & \PP_{12} & \dots  & \PP_{1J} \\
    \PP_{21} & \Id_{22} & \dots  & \PP_{2J} \\
    \vdots & \vdots & \ddots & \vdots \\
    \PP_{J1} & \PP_{J2} & \dots  & \Id_{JJ}
\end{bmatrix} 
\begin{bmatrix}
    u_{1} \\
    u_{2} \\
    \vdots   \\
    u_{J}
\end{bmatrix}  = 
\begin{bmatrix}
    f_{1} \\
    f_{2} \\
    \vdots   \\
    f_{J}
\end{bmatrix}
\end{align}
where the total number of degrees of freedom is $N = N_{1} + N_{2} + ... + N_{J}$.

\subsection{Complexity $\mathcal{O}(N)$} \label{Subsection.Complexity}

\begin{figure}[H]
\centering
\subfloat[$\AA_{1,2}$]{\includegraphics[width=0.32\textwidth]{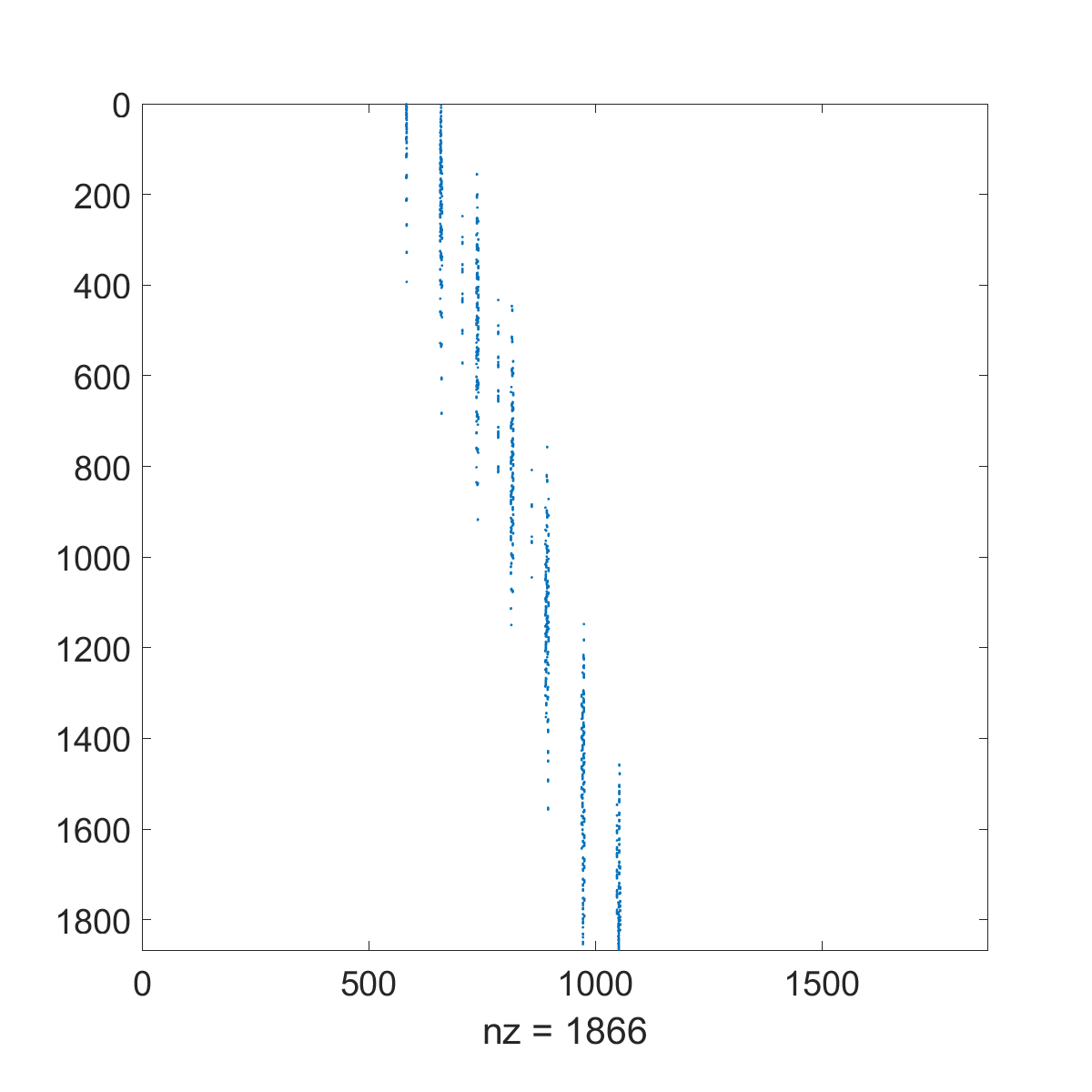}}
\subfloat[$\Delta_{\mathcal{T}_{1}}$]{\includegraphics[width=0.32\textwidth]{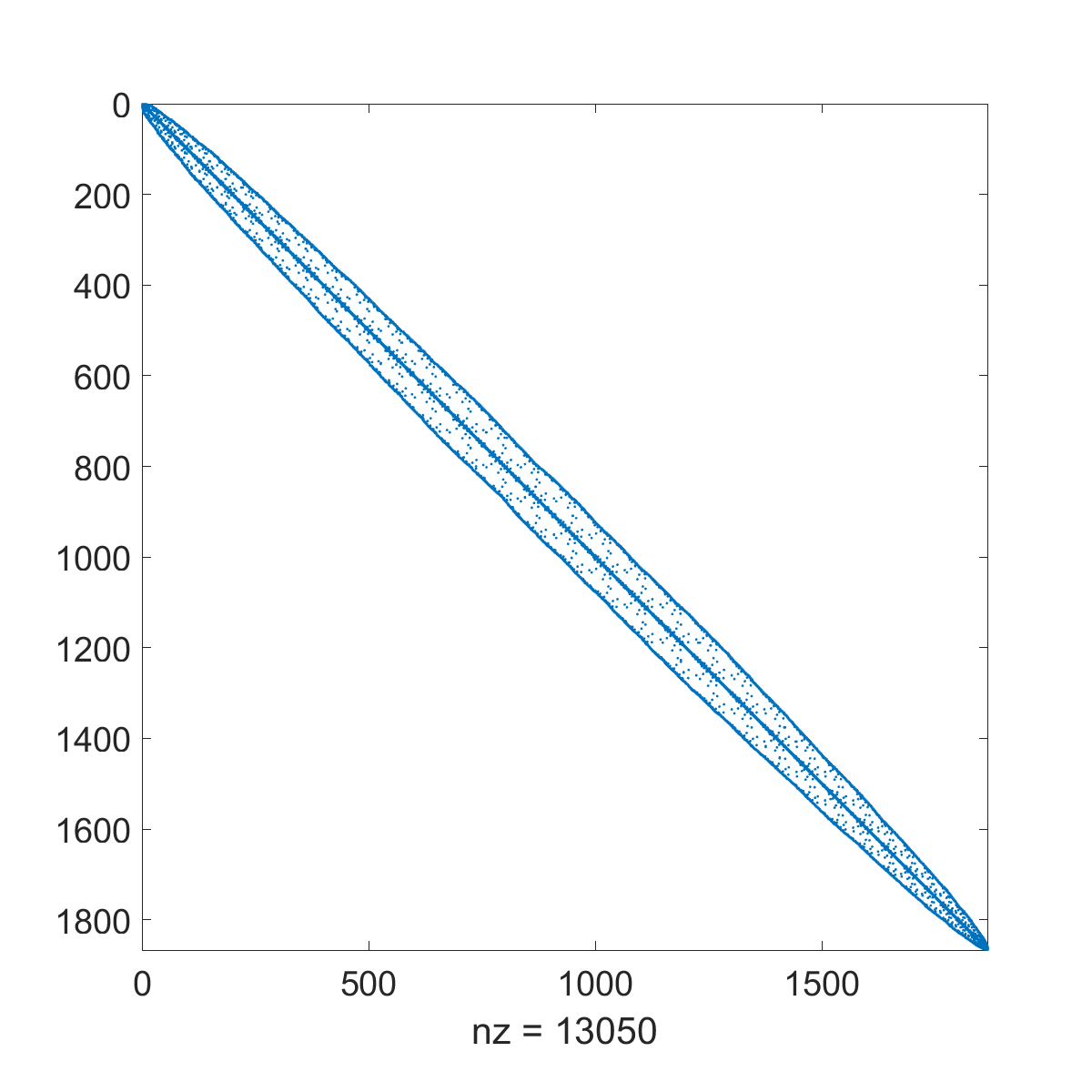}}
\subfloat[CPU time]{\includegraphics[width=0.32\textwidth]{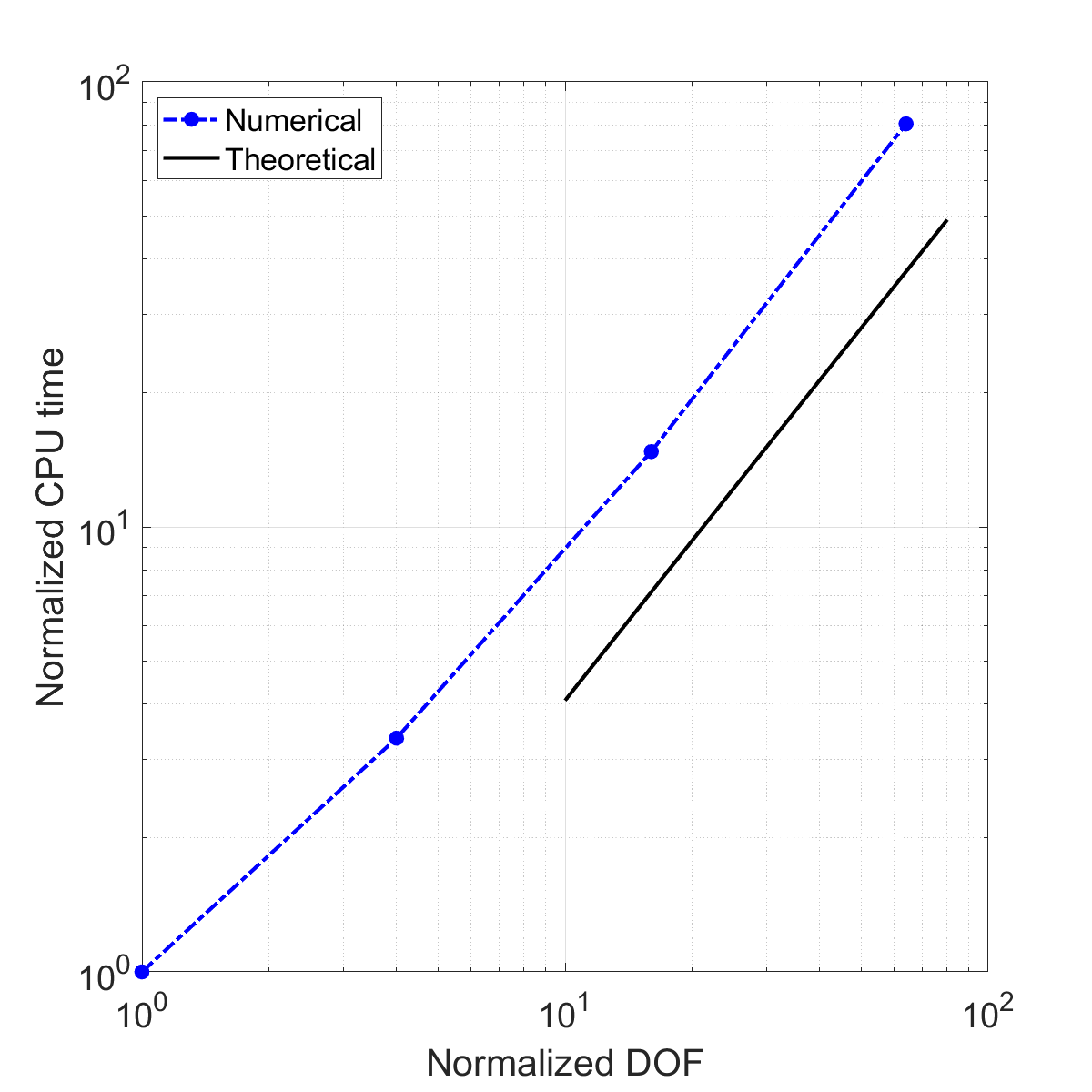}}
\caption{(a) Sparsity pattern of the matrix $\AA_{1,2}$ for the configuration of two spheres from subsection \ref{SubSection.Ex1}. (b) Sparsity pattern of the discrete Laplace-Beltrami operator $\Delta_{\mathcal{T}_{1}}$. (c) Normalized CPU time as a function of normalized DOF for the matrix-vector multiplication associated with system (\ref{Eqn.SystemMatrix}). The theoretical line corresponds to $\mathcal{O}(N)$.}  
\label{Fig:Sparse}
\end{figure}

The standard numerical techniques for BIE, such as Nystrom, collocation and Galerkin boundary element methods, lead to fully populated matrices. Similarly, for the OSRCs developed in
\cite{Acosta2015f,Alzubaidi2016}, matrices with fully populated off-diagonal blocks are obtained.
In these cases, high computational costs $\mathcal{O}(N^2)$ and large memory demands $\mathcal{O}(N^2)$ are needed for matrix-vector multiplication in any iterative solver (e.g. fixed-point or GMRES). By contrast, the proposed OSRC leads to $\mathcal{O}(N)$ complexity. This follows from the propagator matrix (\ref{Eqn.MatrixForm}) that defines the governing system (\ref{Eqn.SystemMatrix}). The implementation of the propagator (\ref{Eqn.MatrixForm}) requires the multiplication by the matrix $\AA_{i,j}$ after the inversion of a system governed by $\Id + \Delta_{\mathcal{T}_{j}} / (2 \iota k  \HH)$. The former is a sparse matrix, with exactly one non-zero entry per row, leading $\mathcal{O}(N_{i})$ operations per matrix-vector multiplication. The latter is the solution for the discrete version of a two-dimensional elliptic equation. With proper sorting of the triangulation nodes, this system is both sparse and narrowly banded. Therefore, a direct solver such as LU decomposition requires $\mathcal{O}(N_{j})$ operations. As a consequence, any iterative solver for the system (\ref{Eqn.SystemMatrix}) will require $\mathcal{O}(N)$ operations per iteration. Figure \ref{Fig:Sparse} displays (a) the sparsity pattern of the matrix $\AA_{1,2}$ for the configuration of two spheres from subsection \ref{SubSection.Ex1}, (b) the sparsity pattern of the discrete Laplace-Beltrami operator $\Delta_{\mathcal{T}_{1}}$, and (c) CPU time for the application of the propagator (\ref{Eqn.MatrixForm}) as a function of DOF. 

For the numerical results presented in the next section, we apply a fixed-point iteration to solve the system (\ref{Eqn.SystemMatrix}). Due to the convexity of the surfaces $\partial \Omega_{j}$ for all $j=1,2,...,J$, the norm of the propagation operator $\PP_{i,j}$ decreases as the distance between the $i^{\text{th}}$ and $j^{\text{th}}$ obstacles increases. Hence, the system (\ref{Eqn.SystemMatrix}) can be solved using the following fixed-point iteration,
\begin{align} \label{Eqn.FixedPoint}
\begin{bmatrix}
    u_{1} \\
    u_{2} \\
    \vdots   \\
    u_{J}
\end{bmatrix}^{(m+1)}  = 
\begin{bmatrix}
    f_{1} \\
    f_{2} \\
    \vdots   \\
    f_{J}
\end{bmatrix} - \begin{bmatrix}
    \textbf{0}_{11} & \PP_{12} & \dots  & \PP_{1J} \\
    \PP_{21} & \textbf{0}_{22} & \dots  & \PP_{2J} \\
    \vdots & \vdots & \ddots & \vdots \\
    \PP_{J1} & \PP_{J2} & \dots  & \textbf{0}_{JJ}
\end{bmatrix} 
\begin{bmatrix}
    u_{1} \\
    u_{2} \\
    \vdots   \\
    u_{J}
\end{bmatrix}^{(m)} \qquad m=0,1,2,..., M
\end{align}
for vanishing initial guess at $m=0$.

\subsection{Numerical results} \label{Section.Results}

\begin{figure}[H]
\centering
\subfloat[Sphere]{\includegraphics[width=0.32\textwidth]{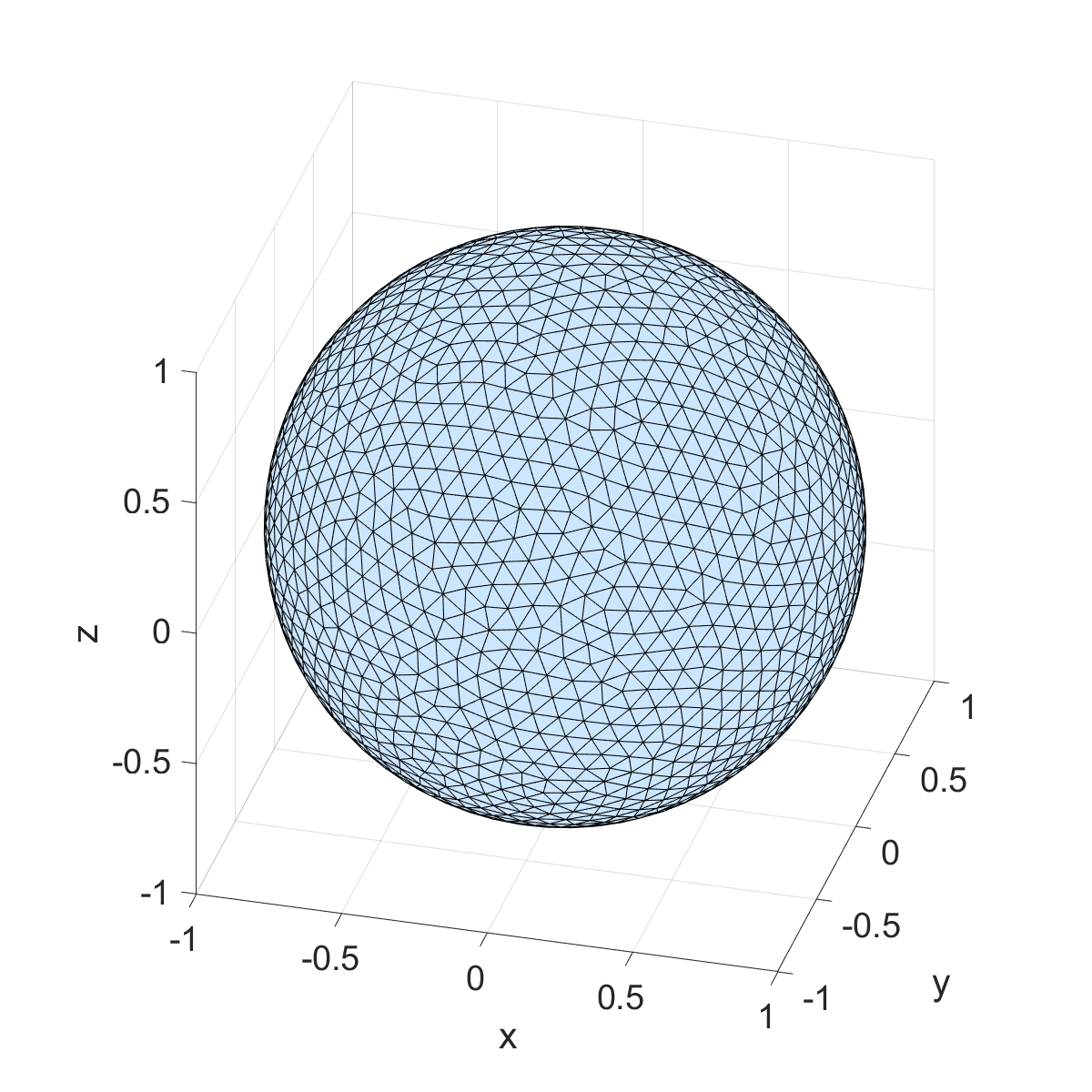}}
\subfloat[Rounded Cube]{\includegraphics[width=0.32\textwidth]{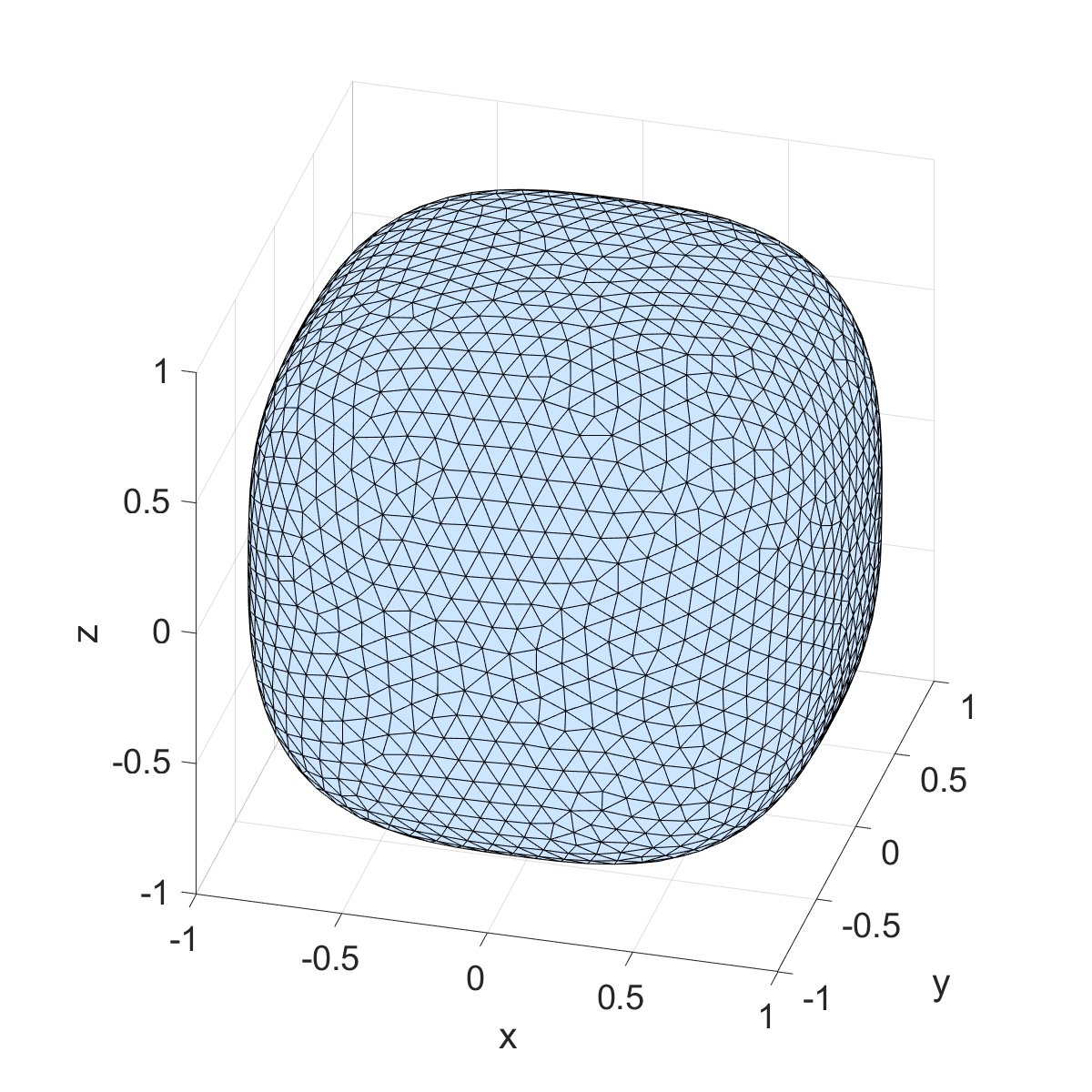}}
\subfloat[Marshmallow]{\includegraphics[width=0.32\textwidth]{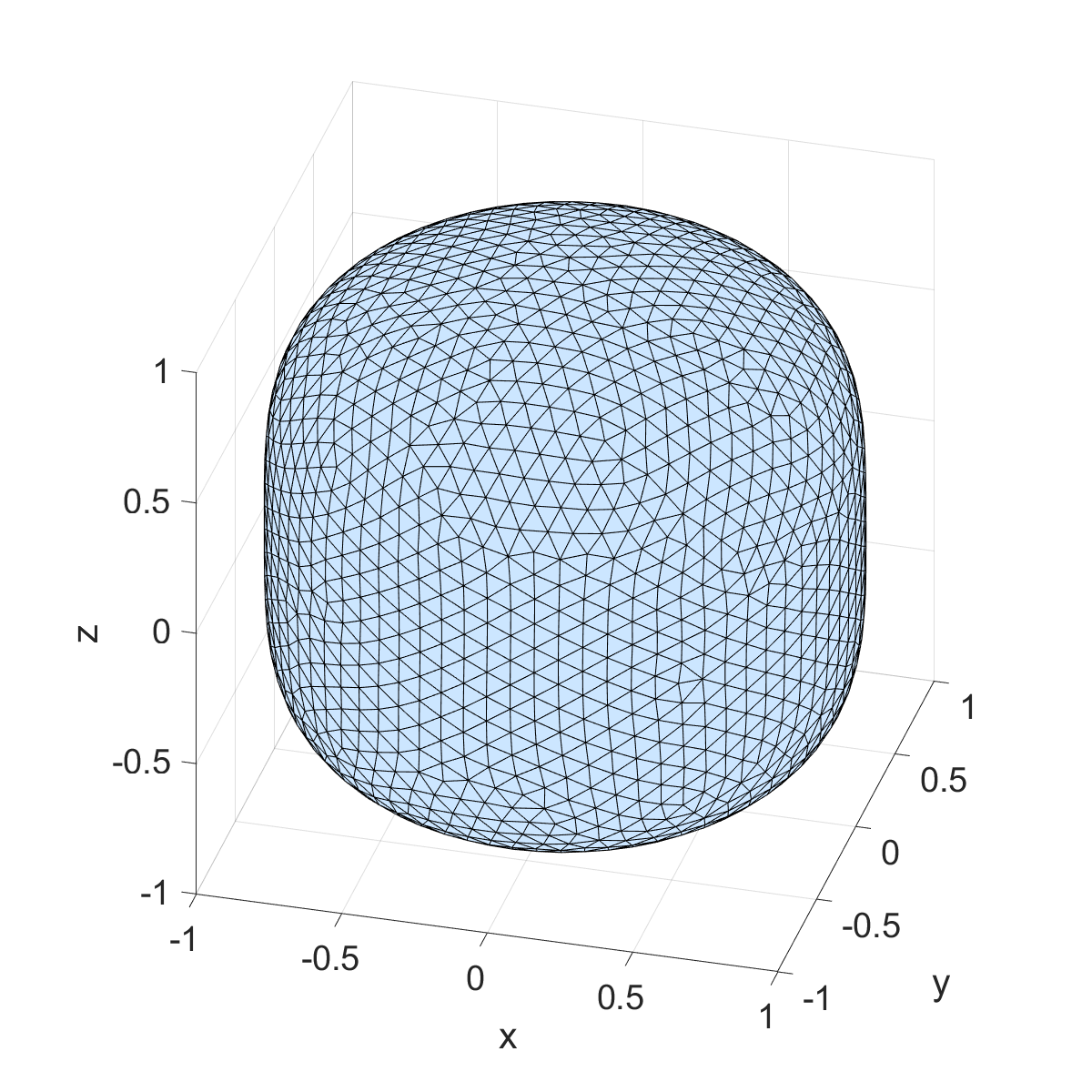}}
\caption{Coarse meshes for the three surfaces employed in the numerical examples.}  
\label{Fig:Meshes}
\end{figure}

\begin{figure}[H]
\centering
\subfloat[Mean Curvature]{\includegraphics[width=0.22\textwidth]{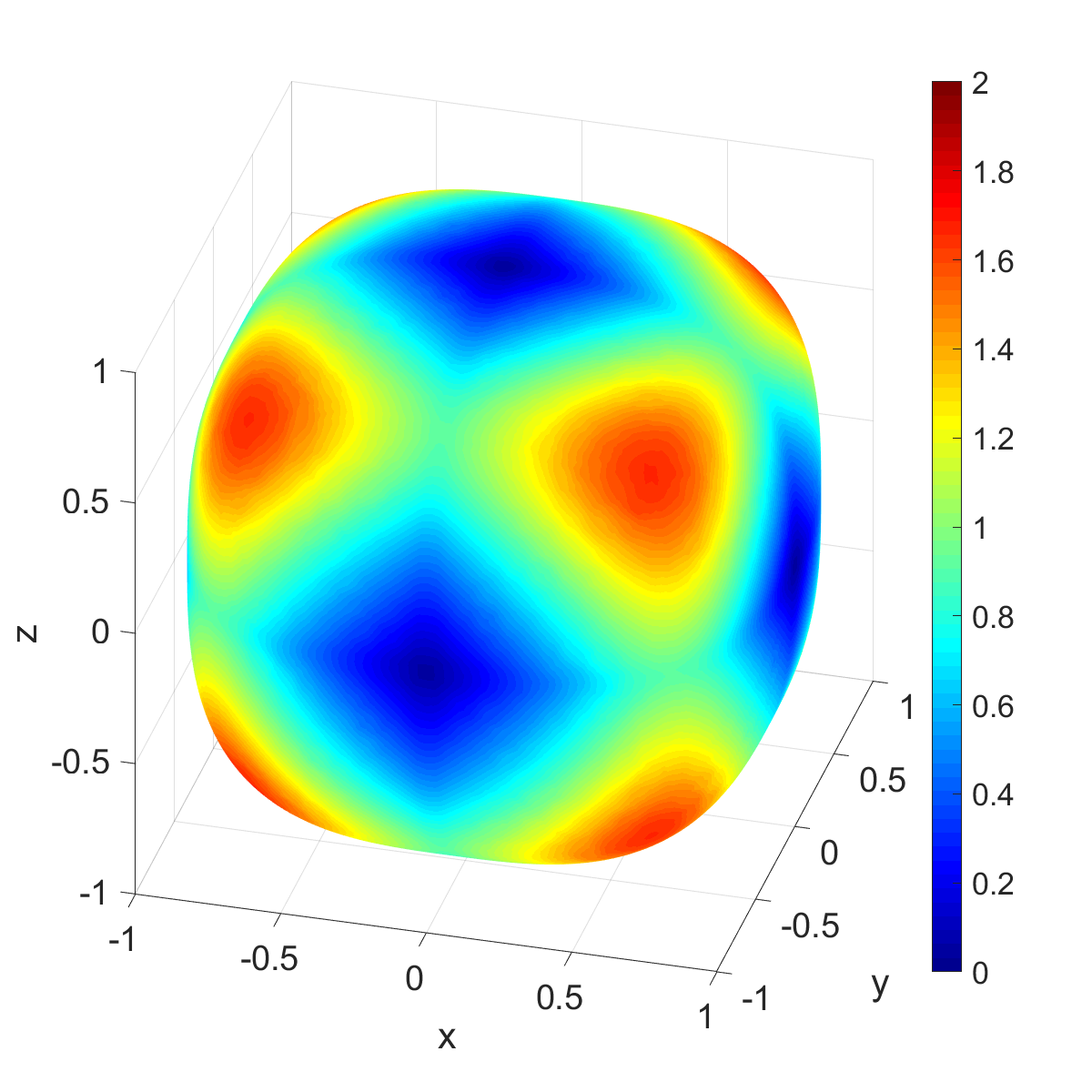}}
\subfloat[Gauss Curvature]{\includegraphics[width=0.22\textwidth]{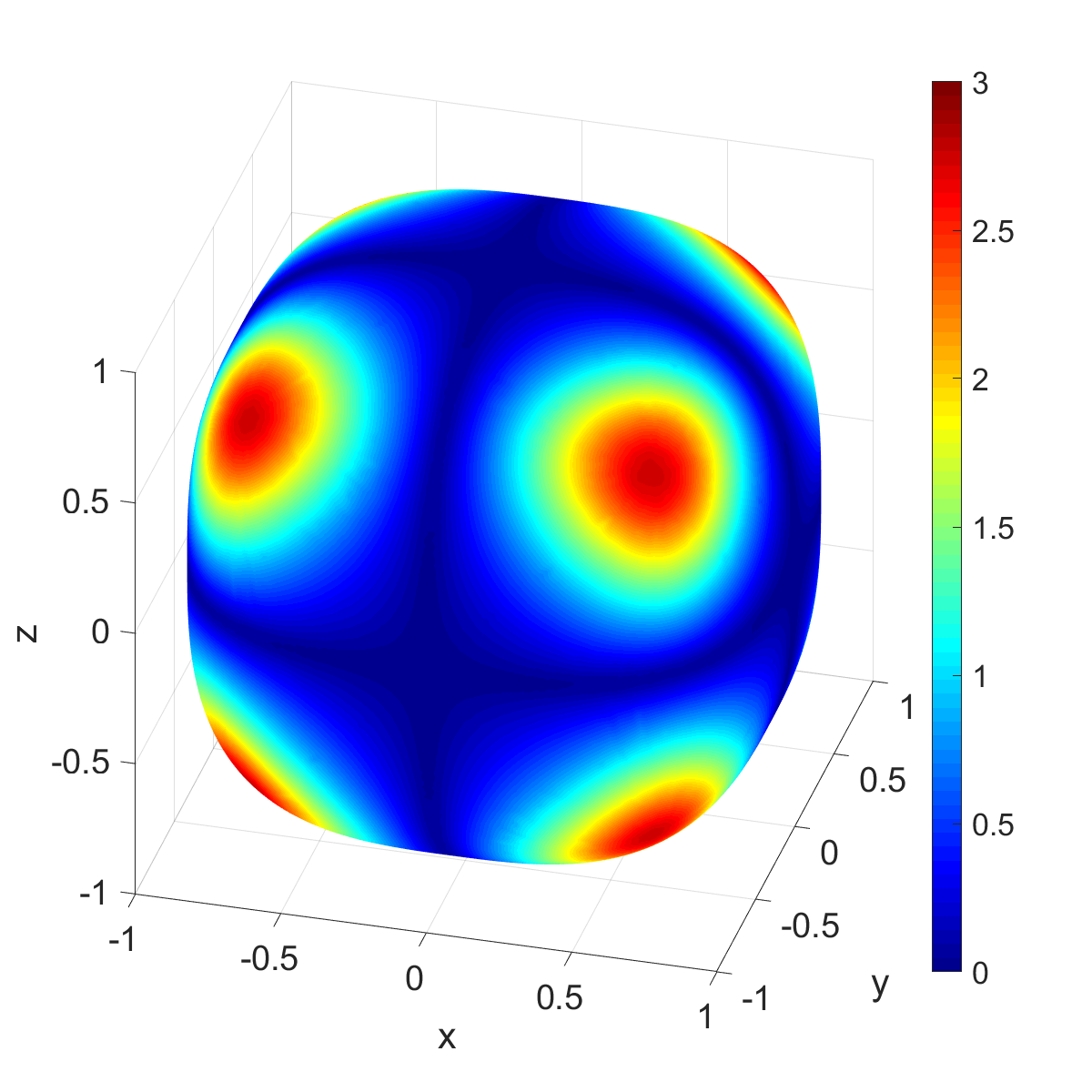}} 
\subfloat[Mean Curvature]{\includegraphics[width=0.22\textwidth]{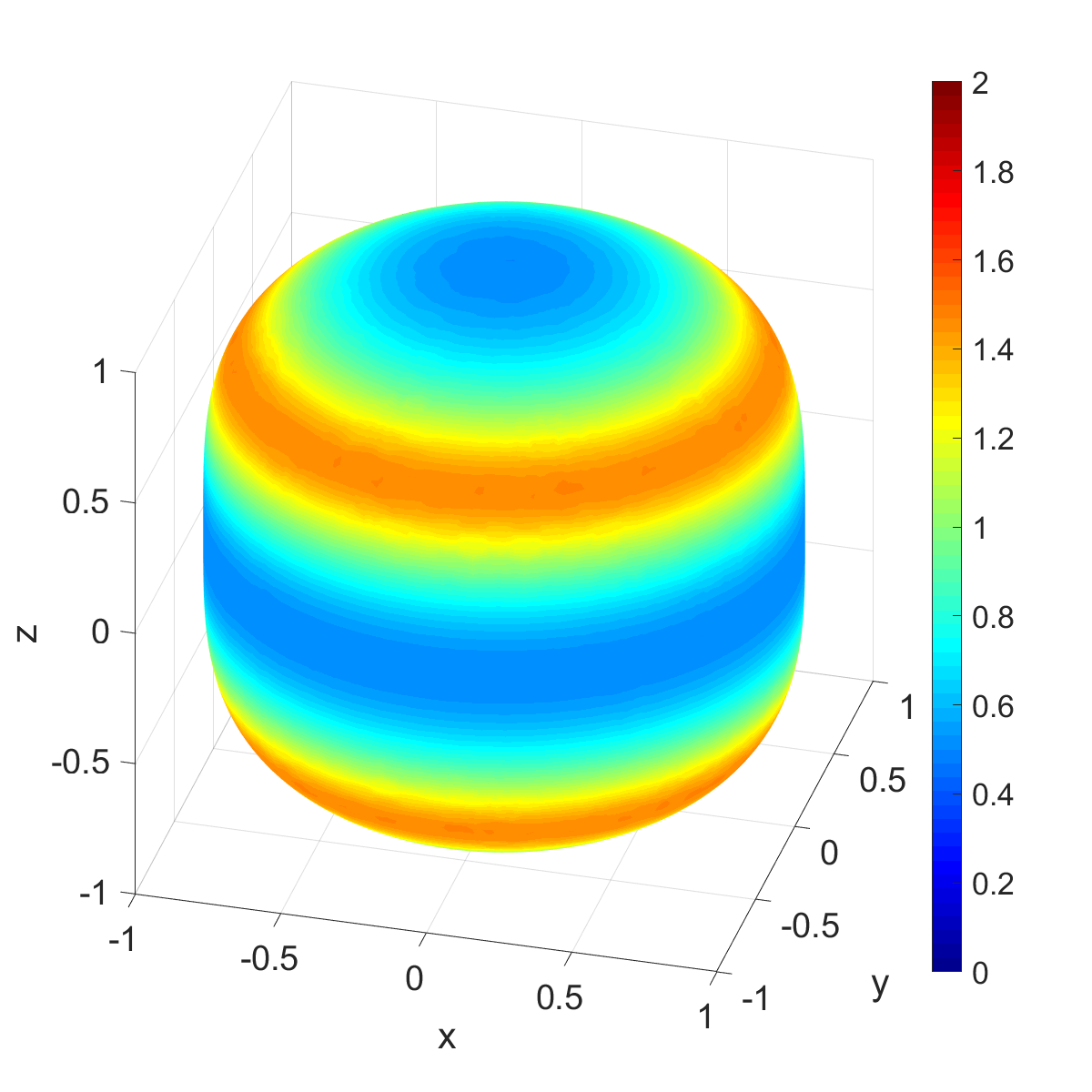}}
\subfloat[Gauss Curvature]{\includegraphics[width=0.22\textwidth]{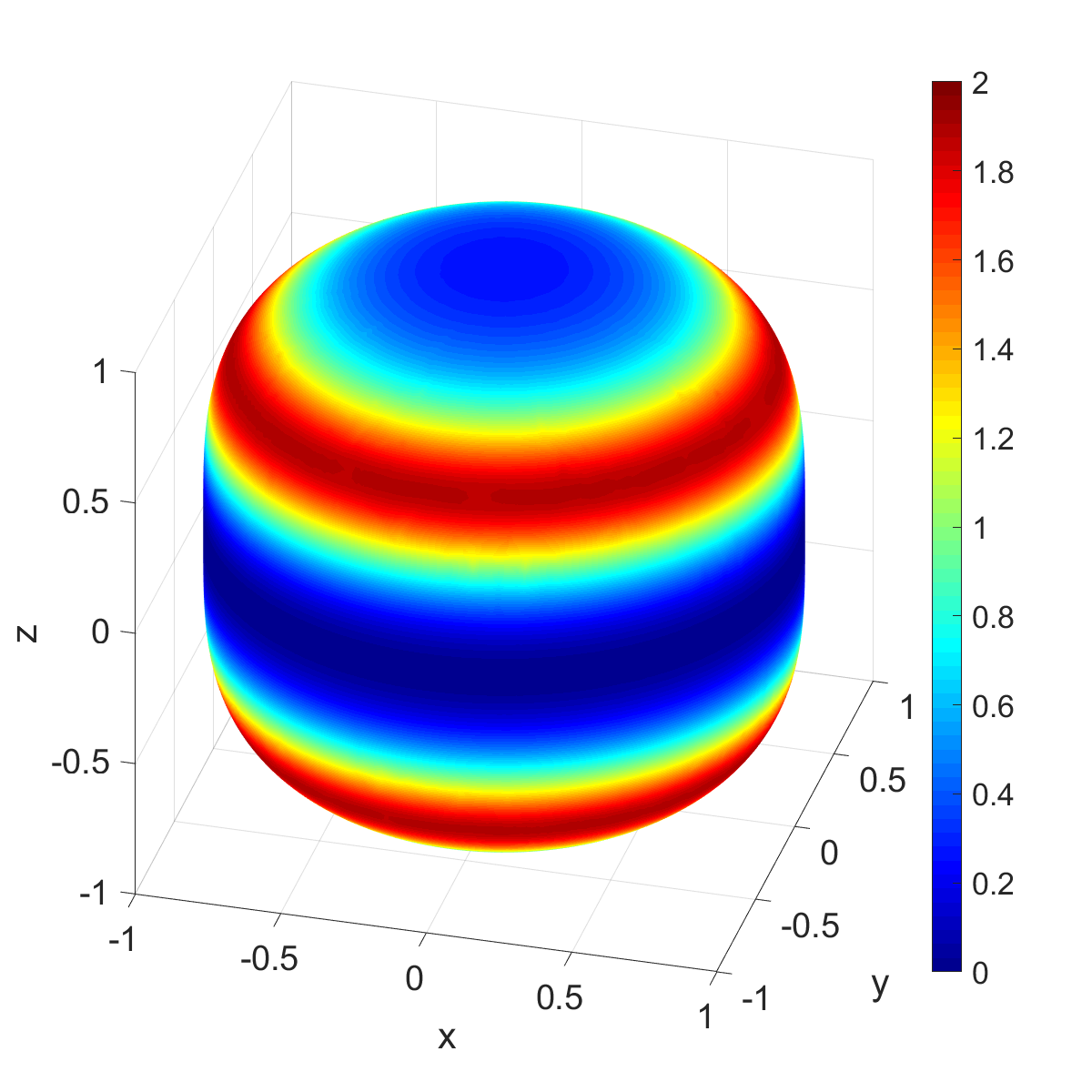}}
\caption{Mean curvature and Gauss curvature for the surfaces employed in the numerical examples.}  
\label{Fig:MeanCurv}
\end{figure}

In this section we present a few numerical results obtained from the implementation of the discrete formulation defined in the previous subsections. To obtain these numerical results, we implemented the fixed-point method (\ref{Eqn.FixedPoint}) with $M=10$ iterations. We consider three convex surfaces for the numerical experiments presented here. These surfaces are the unit sphere, a rounded cube, and a marshmallow--like surface. Their defining equations are as follows,
\begin{subequations} \label{Eqn.SurfacesParametricEqns}
\begin{align} 
&\text{Sphere:} & |x|^2 + |y|^2 + |z|^2 = 1. \\
&\text{Rounded Cube:} & |x|^3 + |y|^3 + |z|^3 = 1. \\
&\text{Marshmallow:} & |x|^2 + |y|^2 + |z|^4 = 1.
\end{align}
\end{subequations}
Coarse triangulations of these three surfaces are shown in Figure \ref{Fig:Meshes}. The discrete mean and Gauss curvatures of the rounded cube and of the marshmallow, defined by (\ref{Eqn.DiscreteMCV})-(\ref{Eqn.DiscreteMC}), are shown in Figure \ref{Fig:MeanCurv}. 

For comparison, we manufacture an exact solution to the boundary values problem (\ref{Eqn.Main}), valid for any geometry of the surfaces in $\partial \Omega$. This is accomplished by defining boundary data $f$ as the boundary trace of a radiating wave field. We consider the field,
\begin{align} \label{Eqn.InternalSource}
F(\xx) = \sum_{j=1}^{J} \Phi(\xx - \cc_{j})
\end{align}
where $\Phi(\zz) = e^{\iota k|\zz|}/(4\pi|\zz|)$ is the outgoing fundamental solution to the Helmholtz equation with frequency $k>0$. Hence, $F$ represents the superposition of $J$ point-sources with respective locations at $\cc_{j}$. If each point $\cc_{j}$ is enclosed by the respective surface $\partial \Omega_{j}$, for $j=1,2,...,J$, then the exact solution to the boundary values problem (\ref{Eqn.Main}) with Dirichlet data $f = F|_{\partial \Omega}$ is given by $F|_{\Omega^{+}}$.

\paragraph{Example 1: Two spheres} \label{SubSection.Ex1}
Now we present some numerical results for the wave radiating problem in the exterior of $J=2$ unit-spheres. These spheres are centered at the follow points:
\begin{align} \label{Eqn.CentersEx1}
\cc_{1} = (2,0,0) \quad \text{and} \quad
\cc_{2} = (-2 , 0 ,0).
\end{align}

Table \ref{table:Error2spheres} displays the $L^2$-norm relative error for the far-field pattern for various wavenumbers and mesh refinements. The DOF approximately quadruples with each mesh refinement, which corresponds to halving the number of points per wavelength. 

\begin{table}[h!]
\centering
\begin{tabular}{r r r r r r r r} \toprule
    {DOF} & {$k=\pi$} & {$k=2\pi$} & {$k=4\pi$} & {$k=8\pi$} & {$k=16\pi$} \\ \midrule
    $3,732$  & $2.92 \times 10^{-4}$ & $8.29 \times 10^{-4}$ & $2.66 \times 10^{-3}$ & $1.35 \times 10^{-2}$ & $4.25 \times 10^{-2}$ \\
    $14,916$ & $8.23 \times 10^{-5}$ & $2.16 \times 10^{-4}$ & $6.72 \times 10^{-4}$ & $3.45 \times 10^{-3}$ & $1.33 \times 10^{-2}$  \\
    $59,652$ & $1.87 \times 10^{-5}$ & $5.67 \times 10^{-5}$ & $1.68 \times 10^{-4}$ & $8.69 \times 10^{-4}$ & $3.42 \times 10^{-3}$  \\ \bottomrule
\end{tabular} 
\caption{$L^2$-norm relative error for the far-field pattern for various wavenumbers and mesh refinements. Multiple-scattering from two spheres centered at the points (\ref{Eqn.CentersEx1}). The DOF approximately quadruples with each mesh refinement.}
\label{table:Error2spheres}
\end{table}

\begin{figure}[H]
\centering
\subfloat[Exact]{\includegraphics[width=0.32\textwidth]{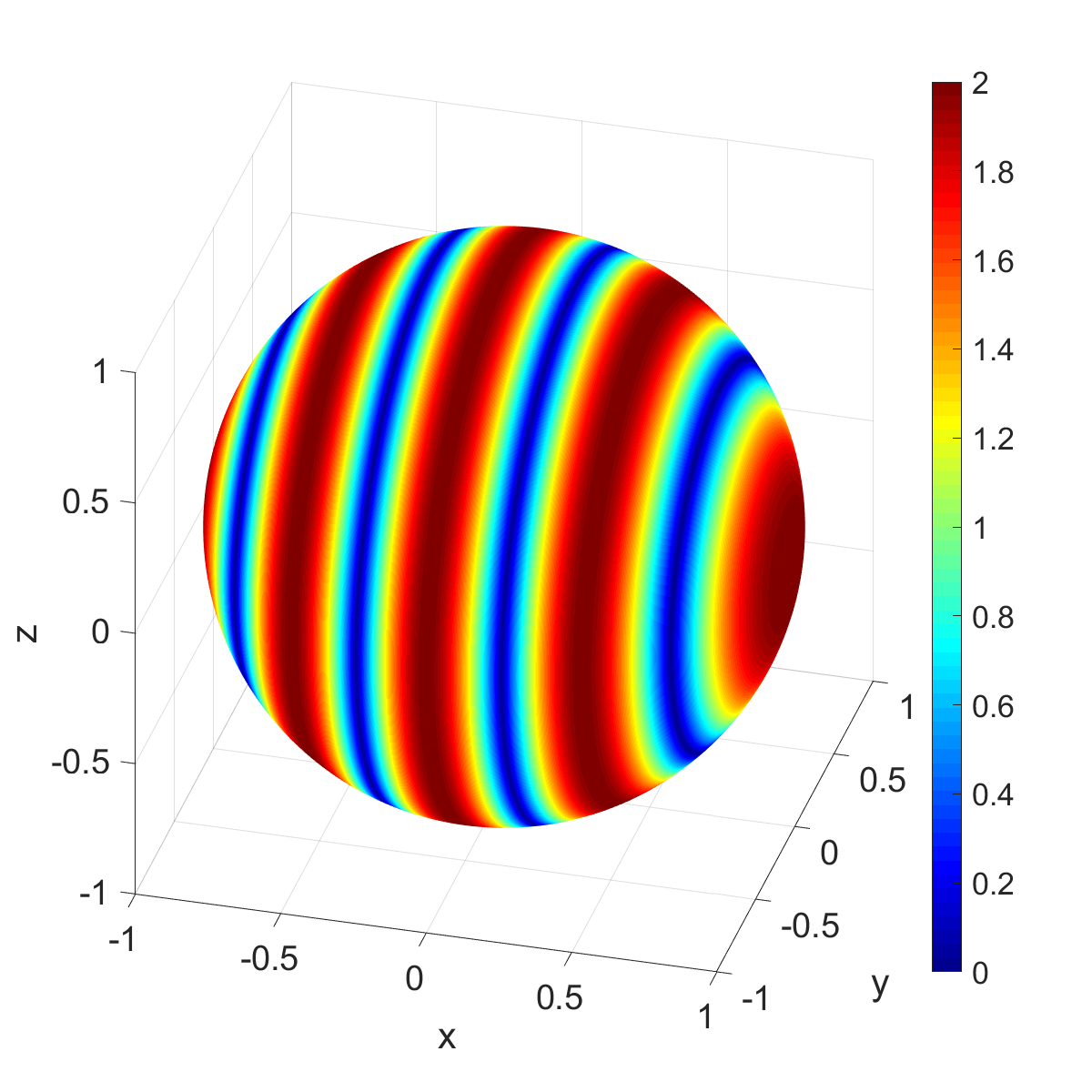}}
\subfloat[OSRC]{\includegraphics[width=0.32\textwidth]{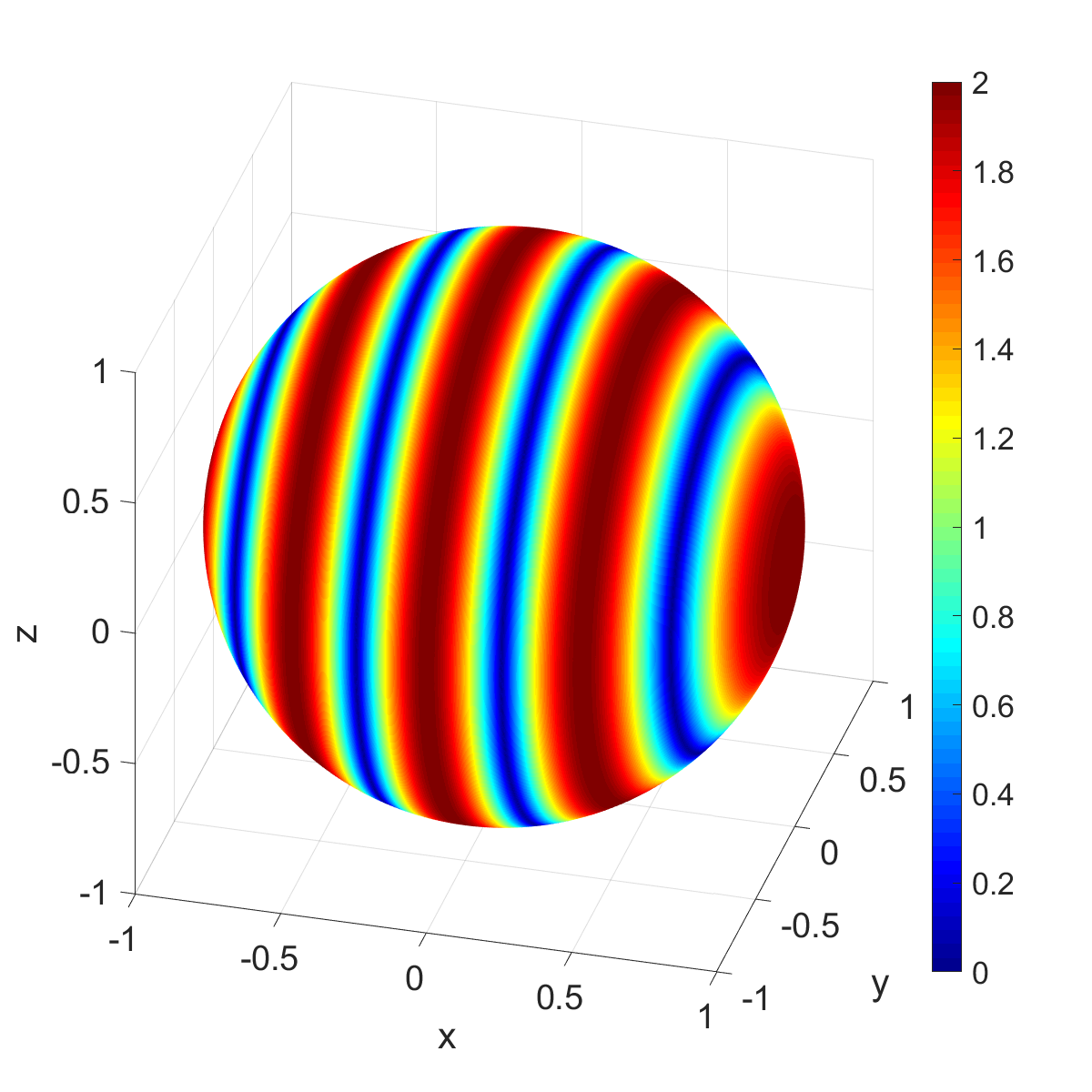}}
\subfloat[Error]{\includegraphics[width=0.32\textwidth]{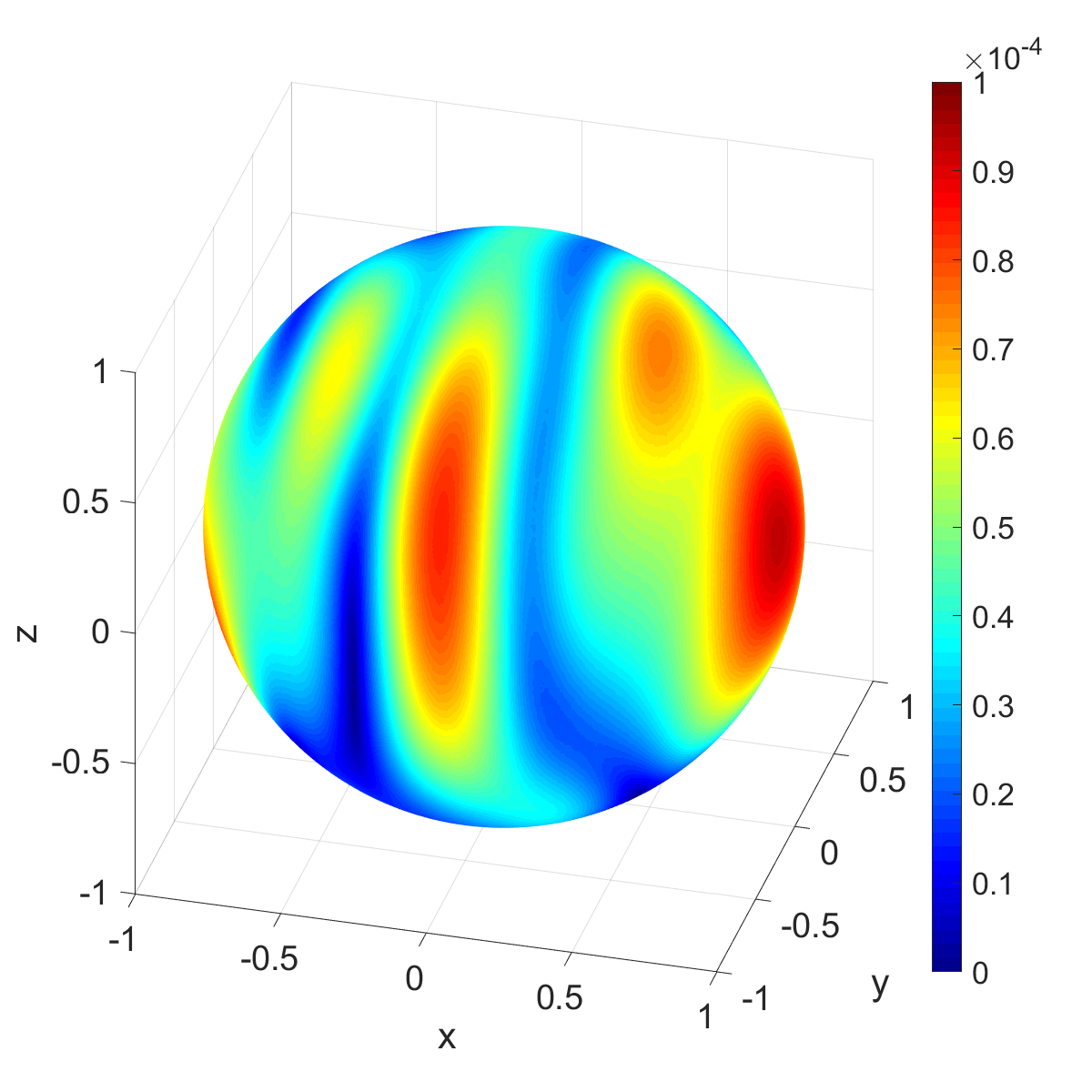}}
\caption{Absolute value of the exact and numerical far-field patterns, and the error profile for the multiple-scattering problem. The wavenumber $k=\pi$.}  
\label{Fig:FFPEx1K1pi}
\end{figure}

\begin{figure}[H]
\centering
\subfloat[Exact]{\includegraphics[width=0.32\textwidth]{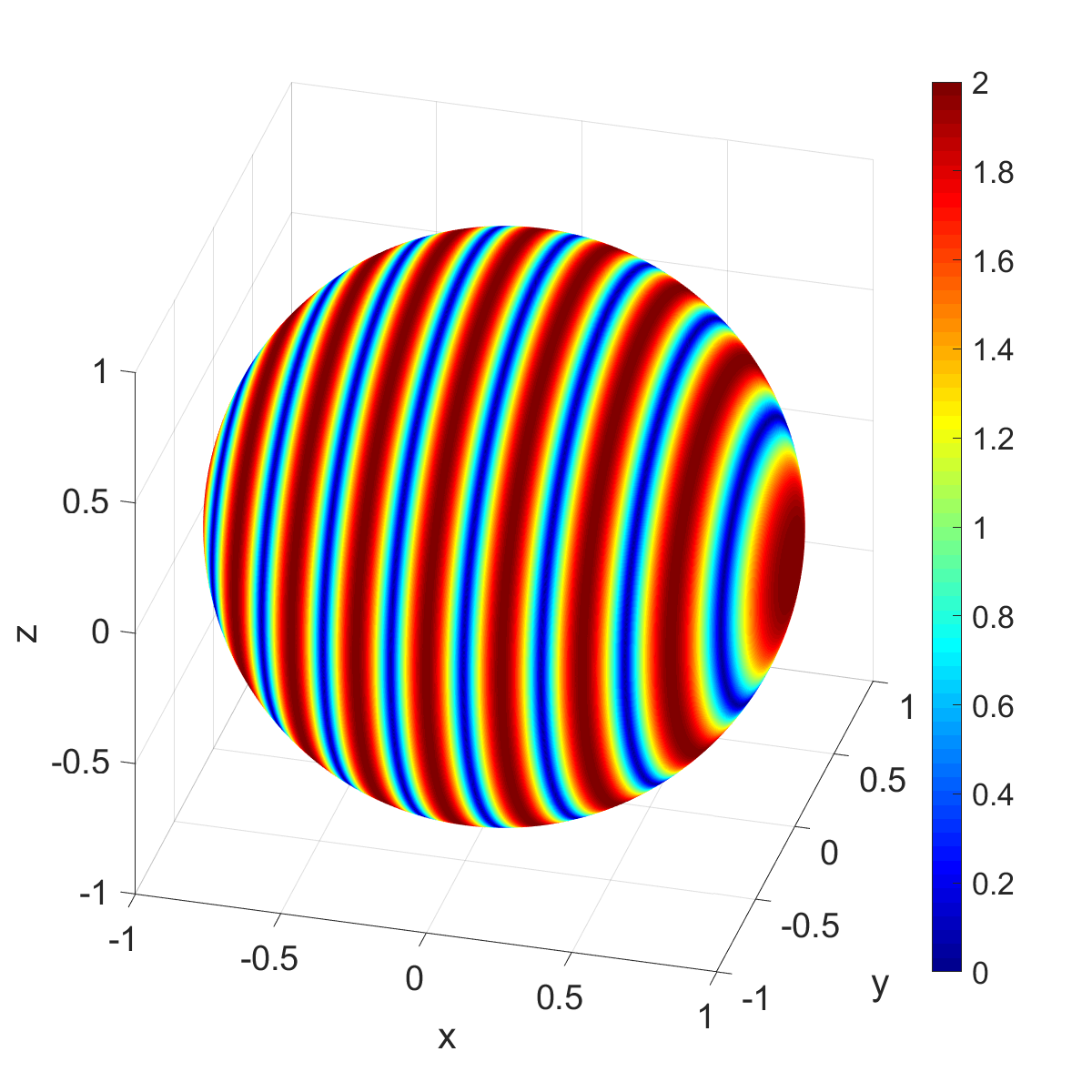}}
\subfloat[OSRC]{\includegraphics[width=0.32\textwidth]{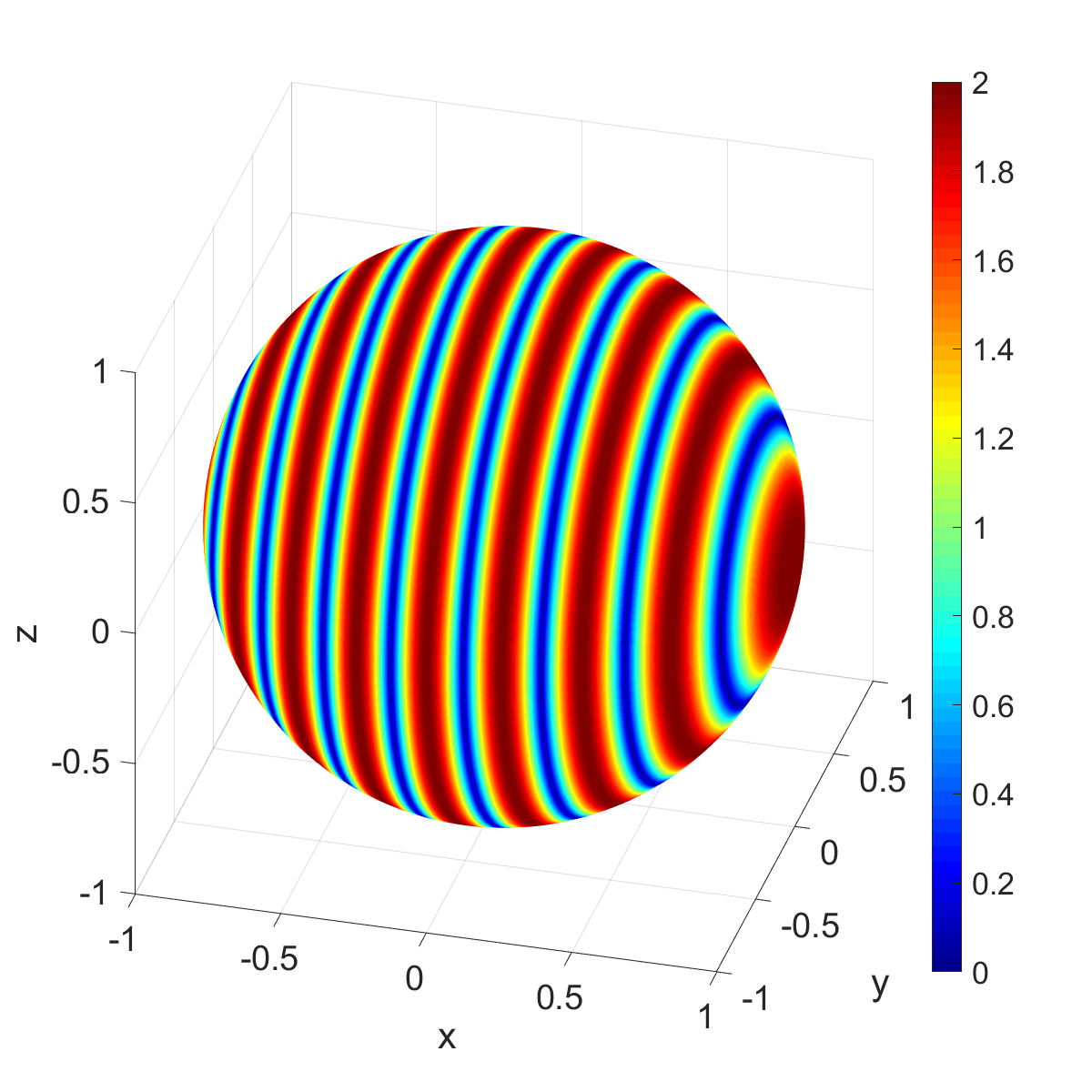}}
\subfloat[Error]{\includegraphics[width=0.32\textwidth]{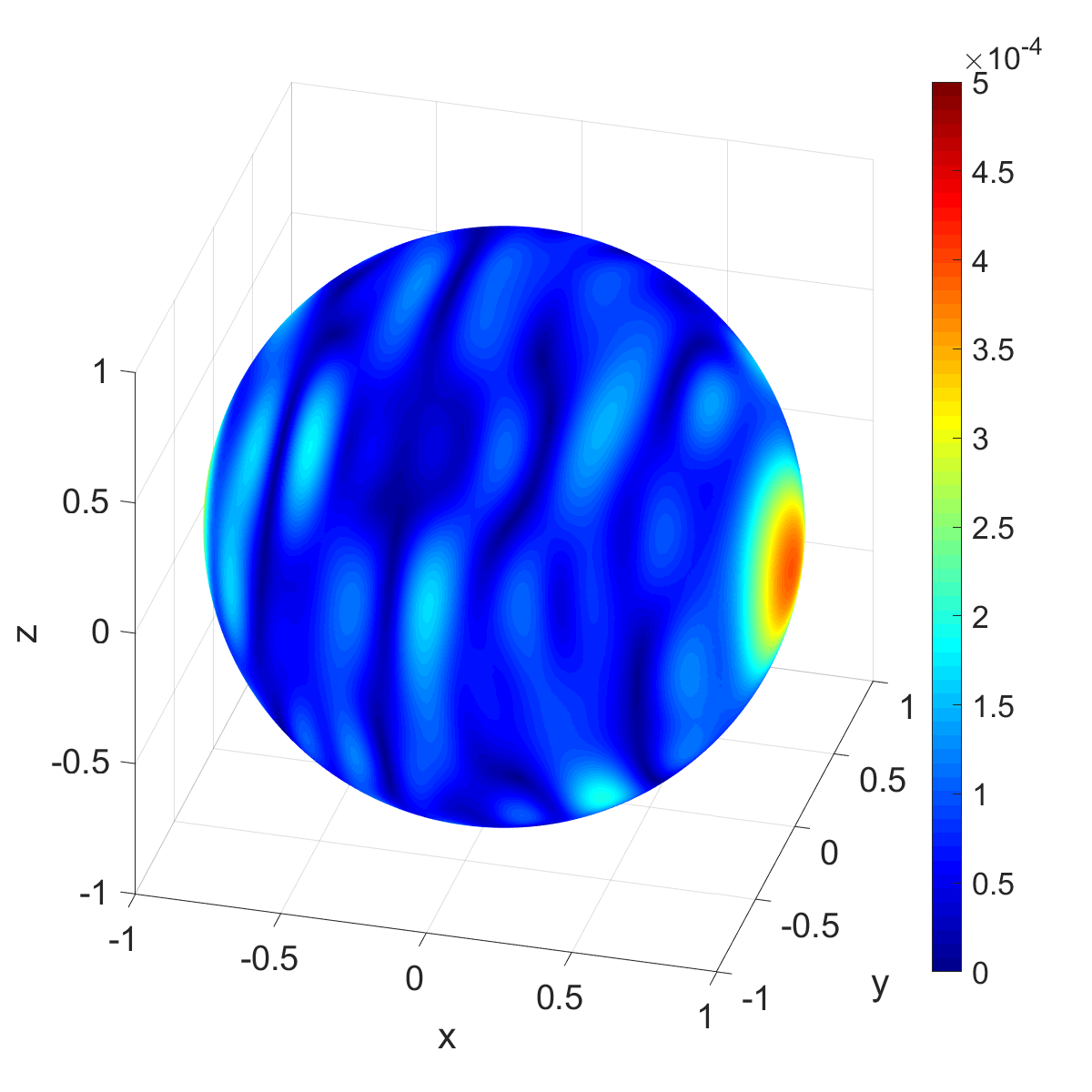}}
\caption{Absolute value of the exact and numerical far-field patterns, and the error profile for the multiple-scattering problem. The wavenumber $k=2 \pi$.}  
\label{Fig:FFPEx1K2pi}
\end{figure}

Figures \ref{Fig:FFPEx1K1pi} and \ref{Fig:FFPEx1K2pi} show the exact and numerical far-field patterns and the error between them, for wavenumbers $k=\pi$ and $k=2 \pi$, respectively.

\begin{figure}[H]
\centering
\includegraphics[width=0.8\textwidth, trim={50 10 50 10}]{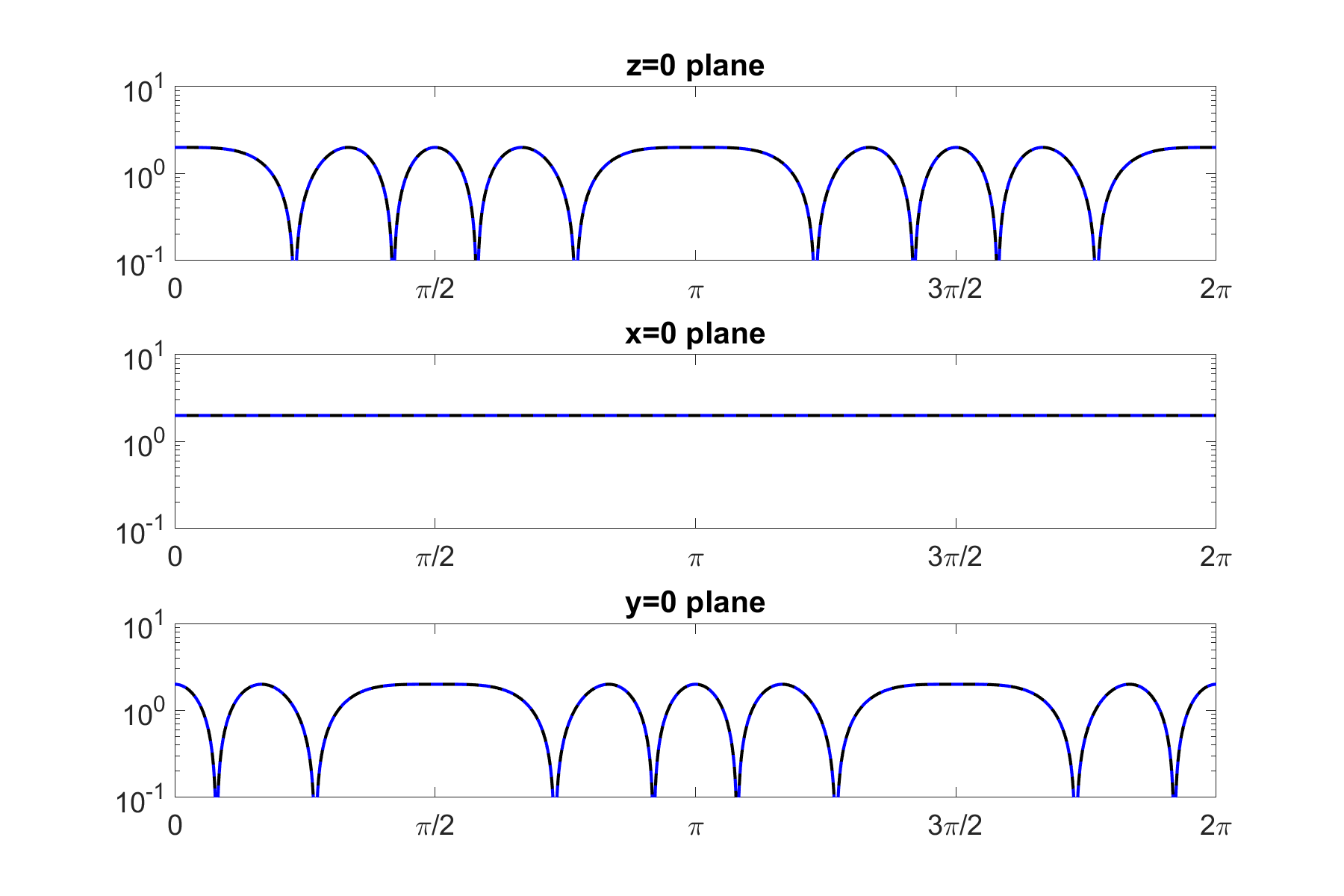}
\caption{Cross-sections of the far-field pattern for the multiple-scattering problem. Black-solid curves are the exact pattern. Blue-dashed curves are the numerical approximation. The wavenumber $k= \pi$.}  
\label{Fig:FFPplanes}
\end{figure}

\begin{figure}[H]
\centering
\includegraphics[width=0.8\textwidth, trim={50 10 50 10}]{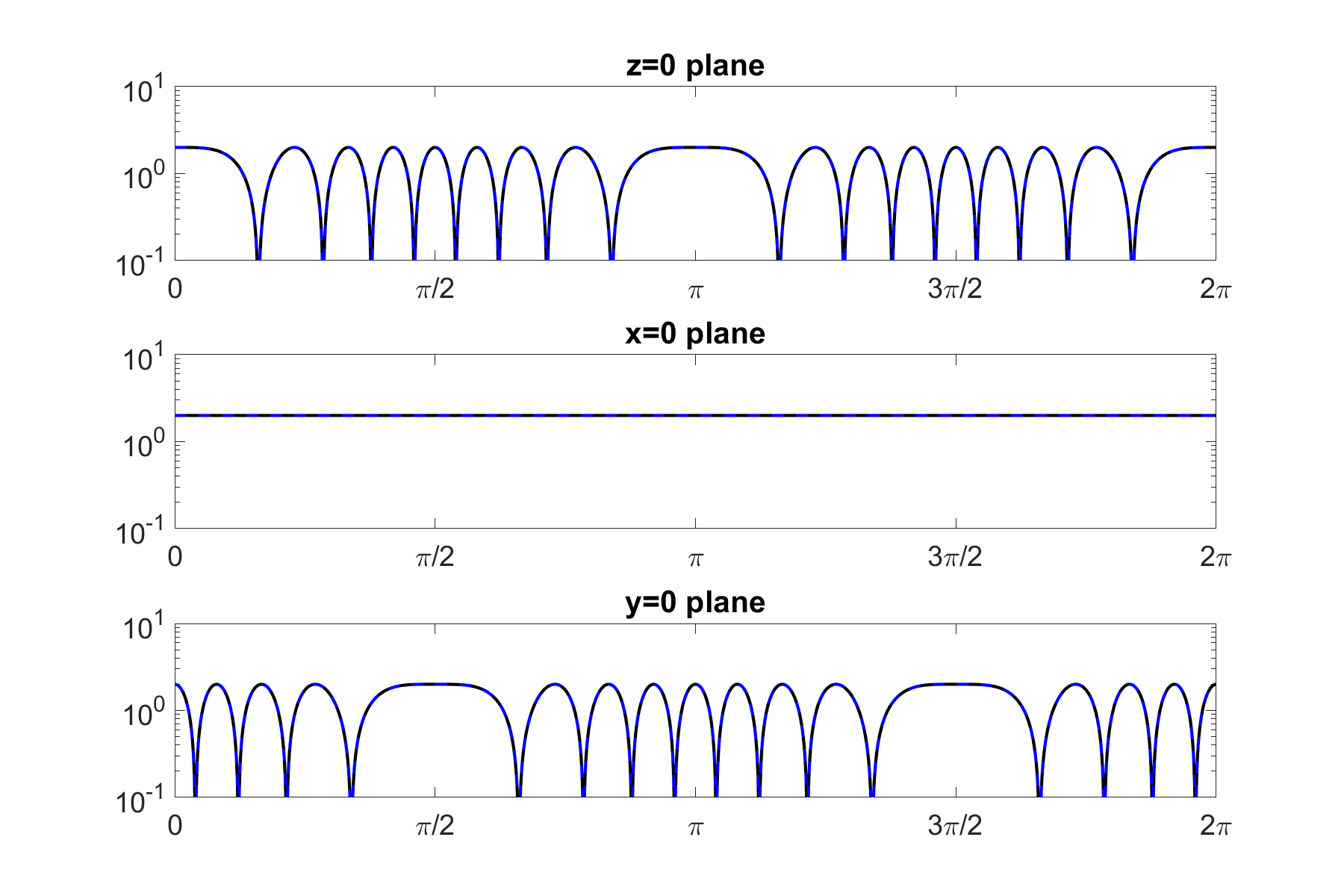}
\caption{Cross-sections of the far-field pattern for the multiple-scattering problem. Black-solid curves are the exact pattern. Blue-dashed curves are the numerical approximation. The wavenumber $k= 2 \pi$.}  
\label{Fig:FFPplanesK2pi}
\end{figure}

For visual comparison, cross-sections of the far-field pattern are shown Figures \ref{Fig:FFPplanes} and \ref{Fig:FFPplanesK2pi} for wavenumbers $k=\pi$ and $k=2 \pi$, respectively.

\paragraph{Example 2: Three different shapes} \label{SubSection.Ex2}

Now we choose $J=3$ surfaces shown in Figure \ref{Fig:Meshes}, translated to the respective points 
\begin{align} \label{Eqn.CentersEx2}
\cc_{1} = 4 \, (1,0,0), \quad
\cc_{2} = 4 \, (\cos \sfrac{2\pi}{3}  , \sin \sfrac{2\pi}{3}  ,0), \quad \text{and} \quad
\cc_{3} = 4 \, (\cos \sfrac{4\pi}{3}  , \sin \sfrac{4\pi}{3}  ,0).
\end{align}

Table \ref{table:Error3shapes} displays the $L^2$-norm relative error for the far-field pattern for various wavenumbers and mesh refinements of these three shapes. The DOF approximately quadruples with each mesh refinement, which corresponds to halving the number of points per wavelength in any tangential direction. 

\begin{table}[h!]
\centering
\begin{tabular}{r r r r r r r r} \toprule
    {DOF} & {$k=\pi$} & {$k=2\pi$} & {$k=4\pi$} & {$k=8\pi$} & {$k=16\pi$} \\ \midrule
    $5,842$  & $5.93 \times 10^{-2}$ & $5.58 \times 10^{-2}$ & $3.96 \times 10^{-2}$ & $4.51 \times 10^{-2}$ & $1.09 \times 10^{-1}$ \\
    $23,350$ & $6.07 \times 10^{-2}$ & $5.91 \times 10^{-2}$ & $4.26 \times 10^{-2}$ & $4.53 \times 10^{-2}$ & $6.41 \times 10^{-2}$  \\
    $93,382$ & $6.12 \times 10^{-2}$ & $5.99 \times 10^{-2}$ & $4.37 \times 10^{-2}$ & $4.57 \times 10^{-2}$ & $6.17 \times 10^{-2}$  \\ \bottomrule
\end{tabular} 
\caption{$L^2$-norm relative error for the far-field pattern for various wavenumbers and mesh refinements. Multiple scattering from three surfaces (\ref{Eqn.SurfacesParametricEqns}) centered at the points (\ref{Eqn.CentersEx2}) The DOF approximately quadruples with each mesh refinement.}
\label{table:Error3shapes}
\end{table}

\begin{figure}[H]
\centering
\subfloat[Exact]{\includegraphics[width=0.32\textwidth]{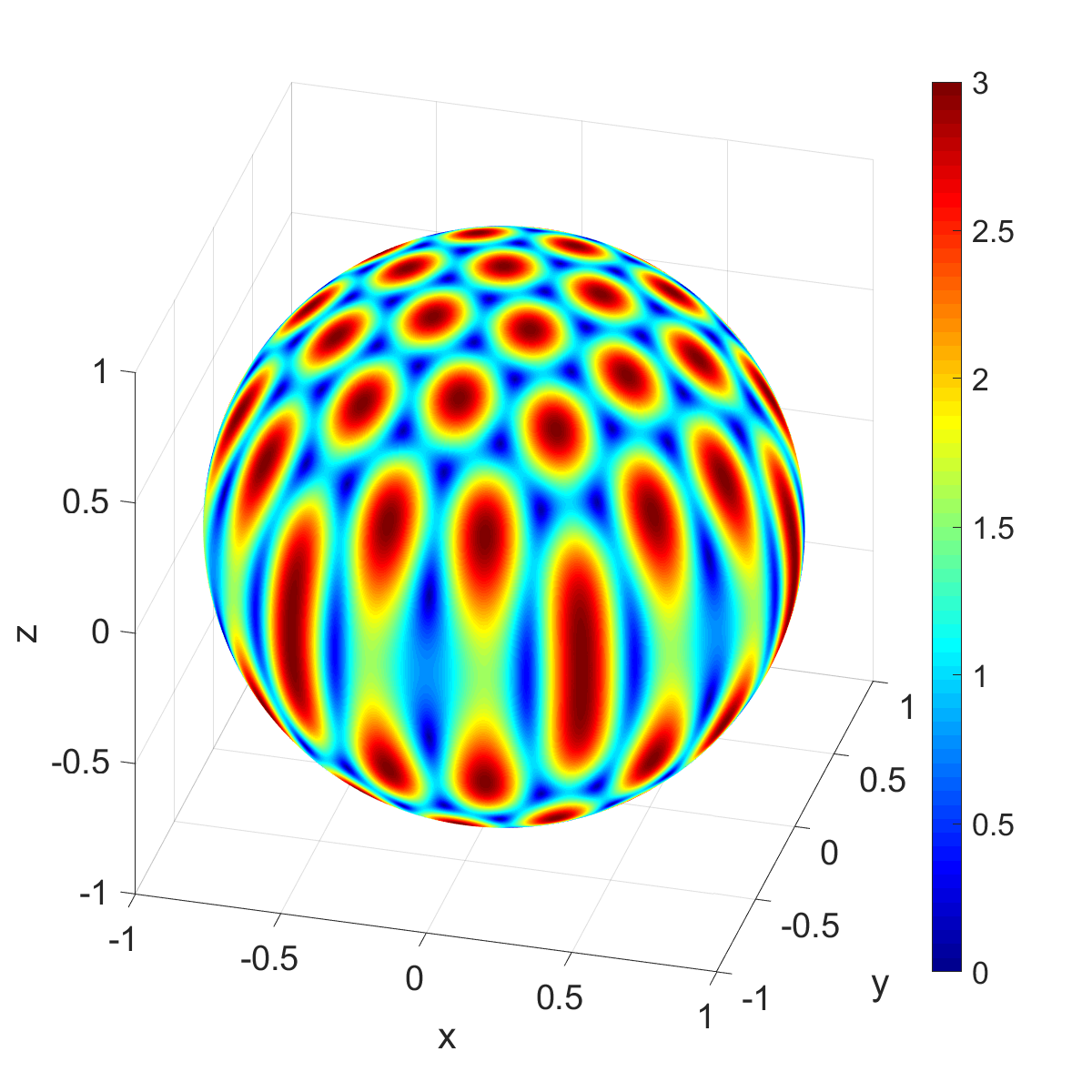}}
\subfloat[OSRC]{\includegraphics[width=0.32\textwidth]{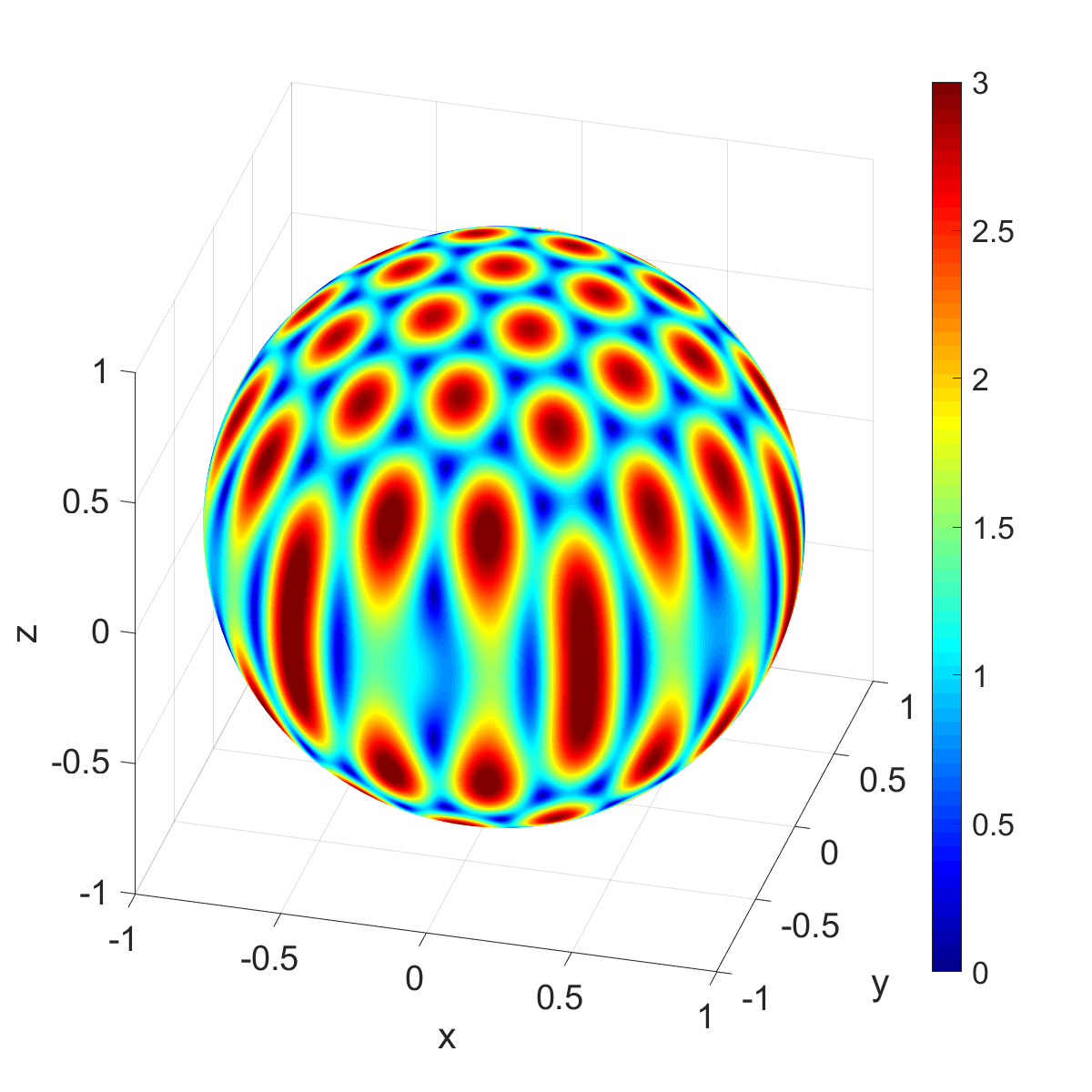}}
\subfloat[Error]{\includegraphics[width=0.32\textwidth]{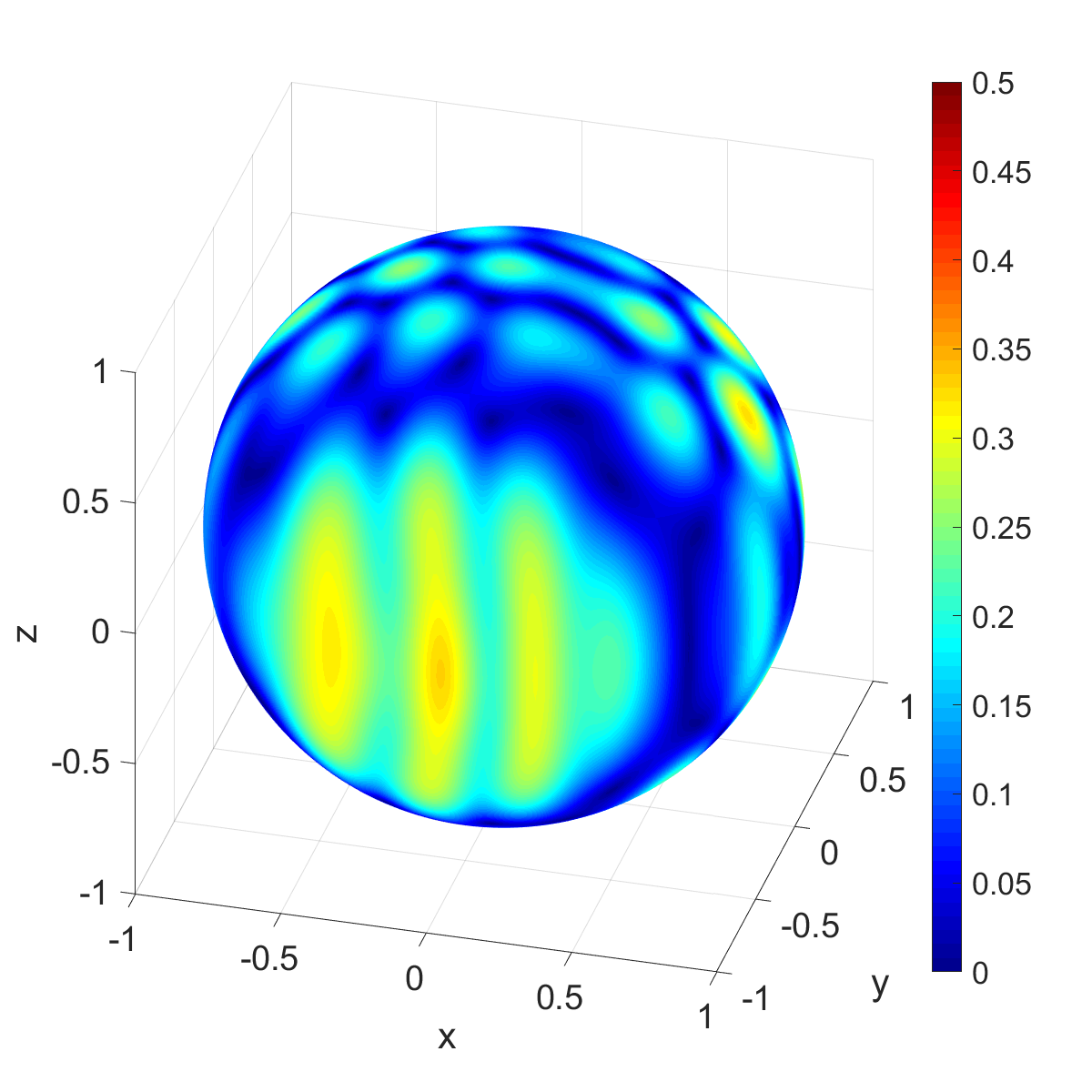}}
\caption{Absolute value of the exact and numerical far-field patterns, and the error profile for the multiple-scattering problem.  The wavenumber $k=\pi$.}  
\label{Fig:FFPEx2K1pi}
\end{figure}

\begin{figure}[H]
\centering
\subfloat[Exact]{\includegraphics[width=0.32\textwidth]{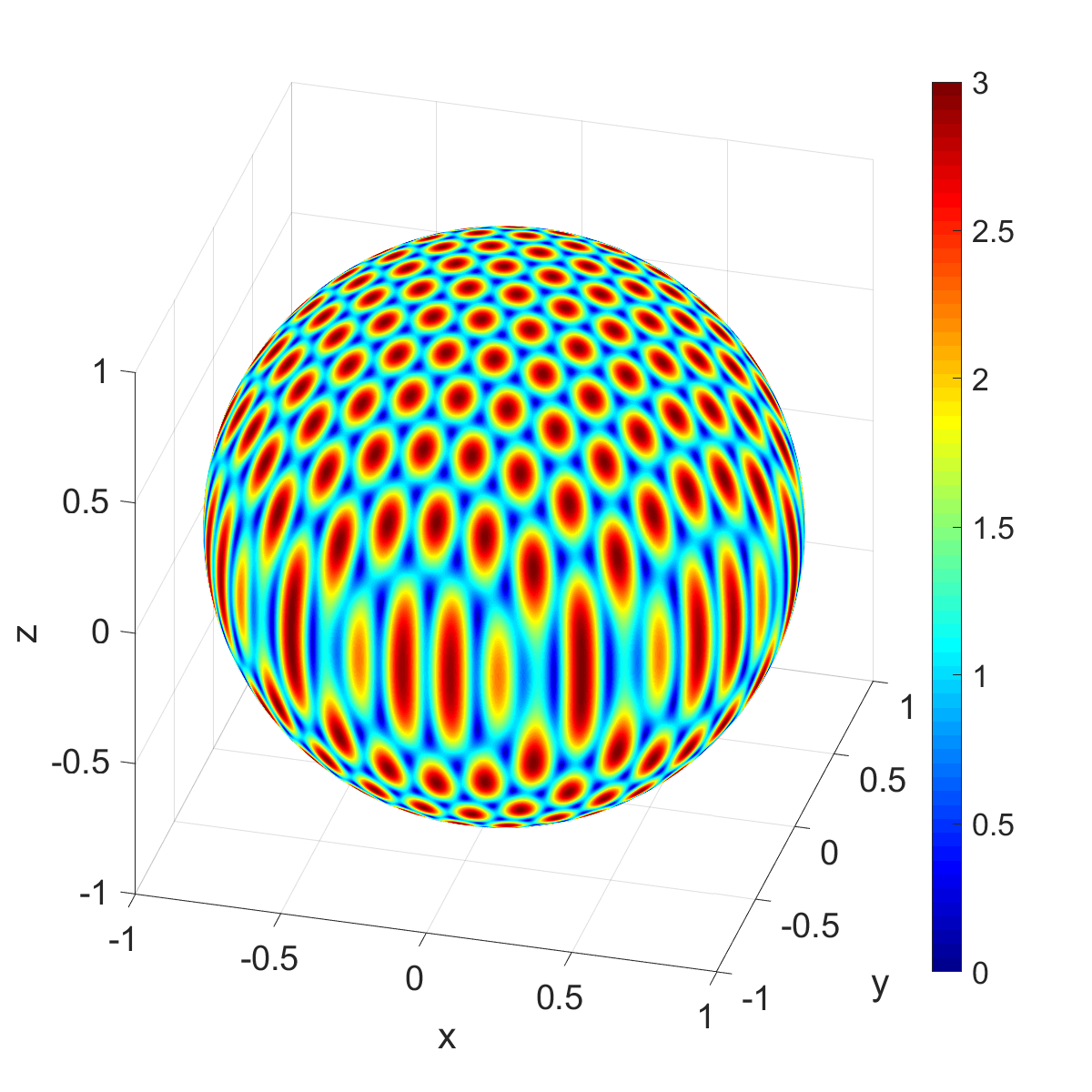}}
\subfloat[OSRC]{\includegraphics[width=0.32\textwidth]{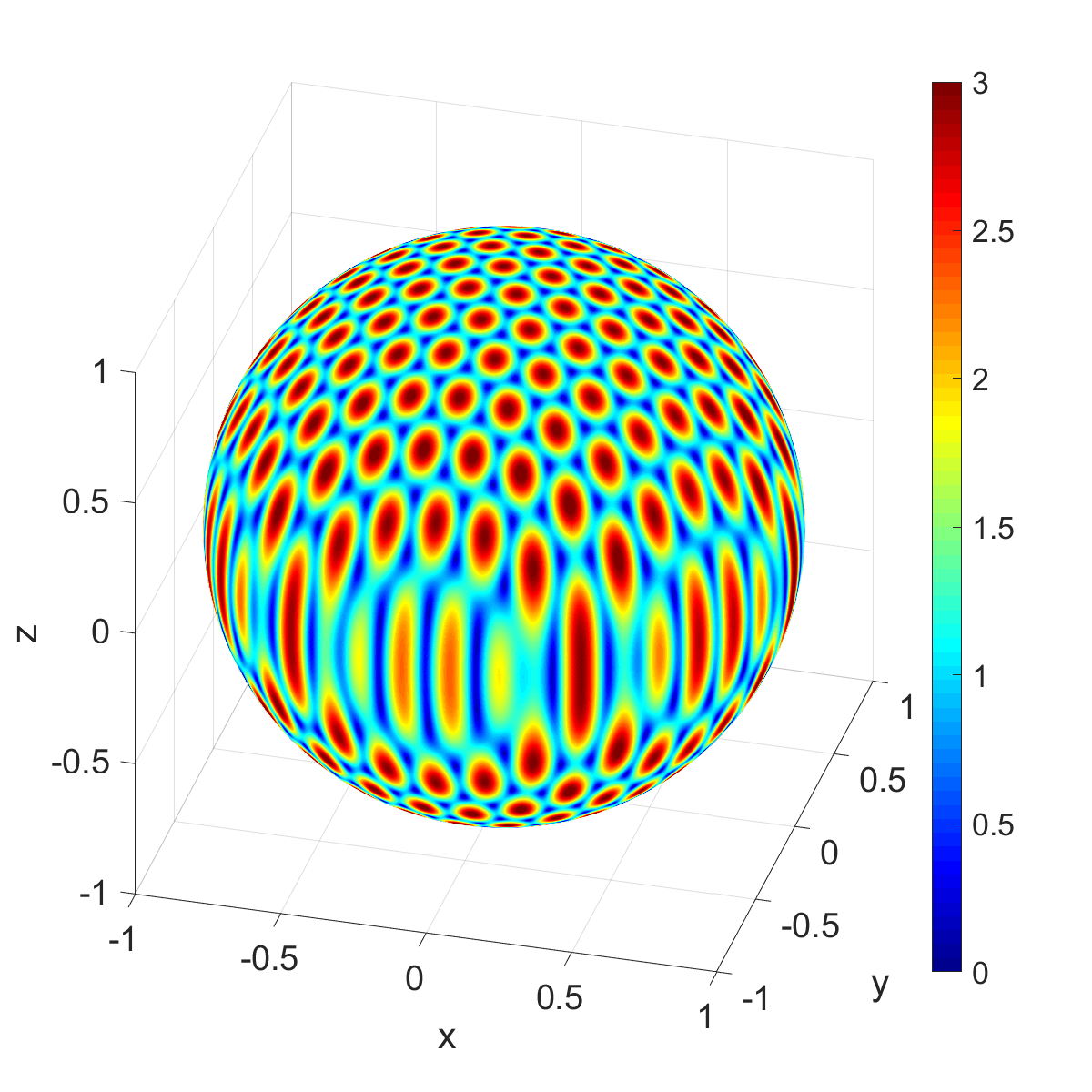}}
\subfloat[Error]{\includegraphics[width=0.32\textwidth]{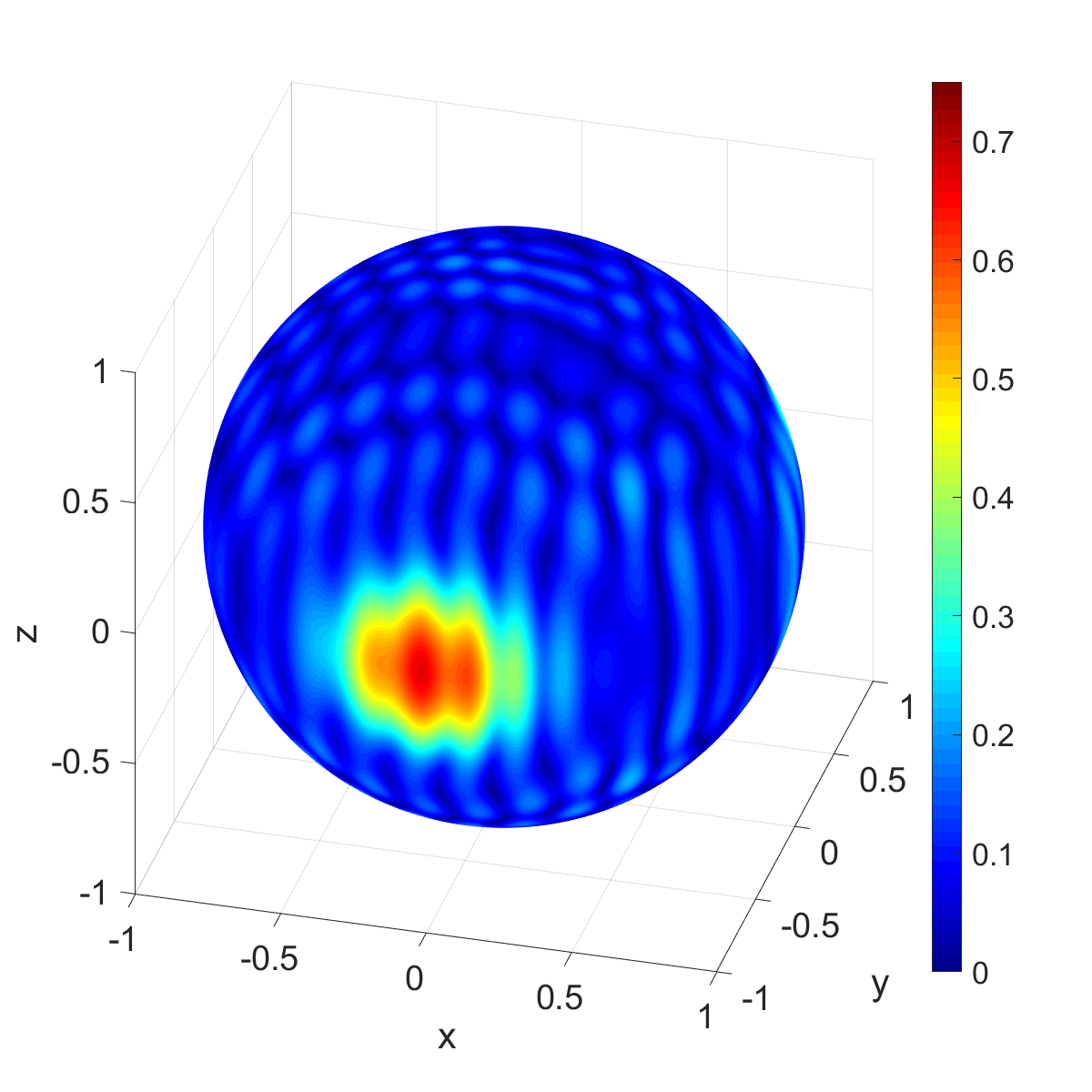}}
\caption{Absolute value of the exact and numerical far-field patterns, and the error profile for the multiple-scattering problem. The wavenumber $k=2 \pi$.}  
\label{Fig:FFPEx2K2pi}
\end{figure}

\begin{figure}[H]
\centering
\includegraphics[width=0.8\textwidth, trim={50 10 50 10}]{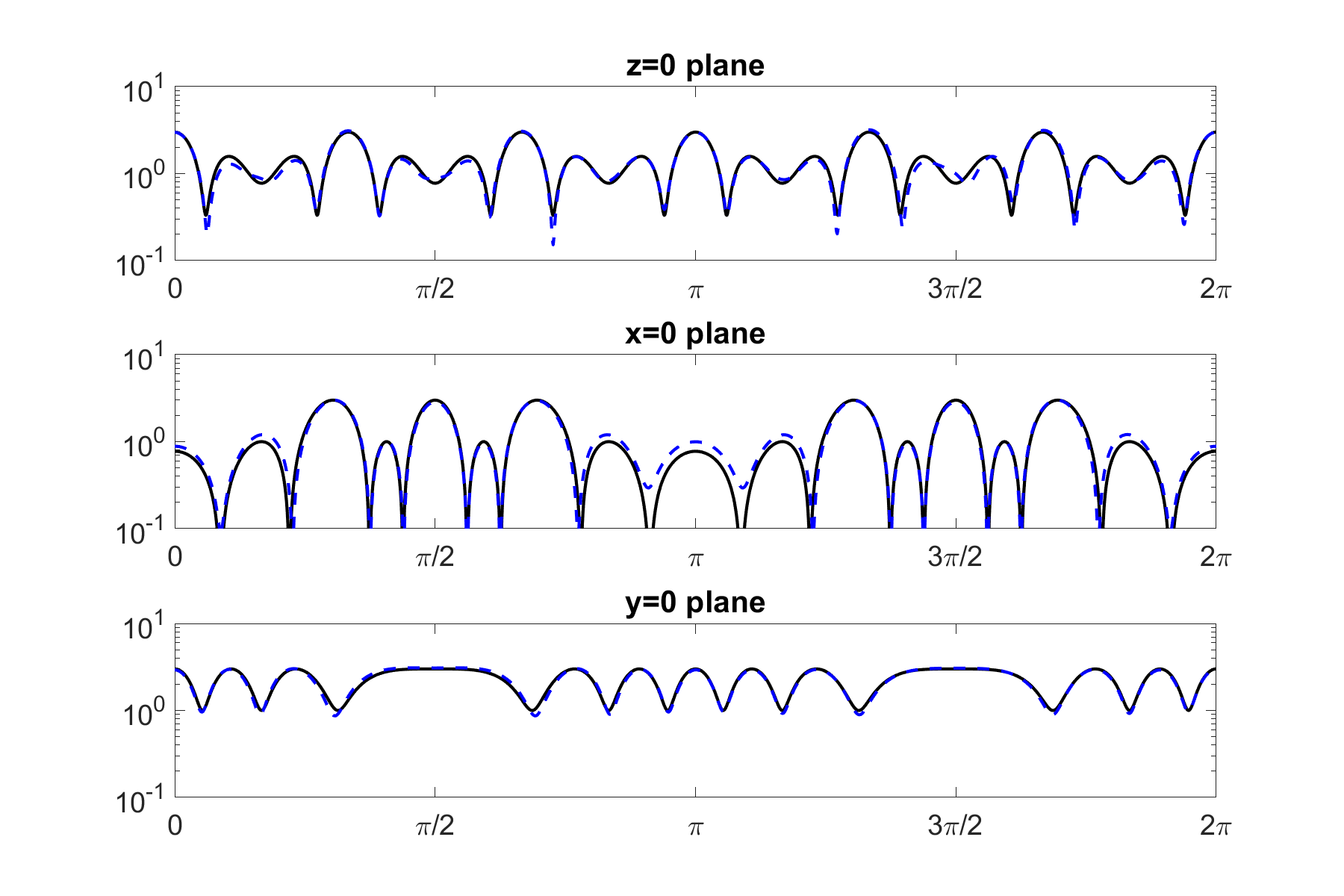}
\caption{Cross-sections of the far-field pattern for the multiple-scattering problem. Black-solid curves are the exact pattern. Blue-dashed curves are the numerical approximation. The wavenumber $k= \pi$.}  
\label{Fig:FFPplanesEx2}
\end{figure}

\begin{figure}[H]
\centering
\includegraphics[width=0.8\textwidth, trim={50 10 50 10}]{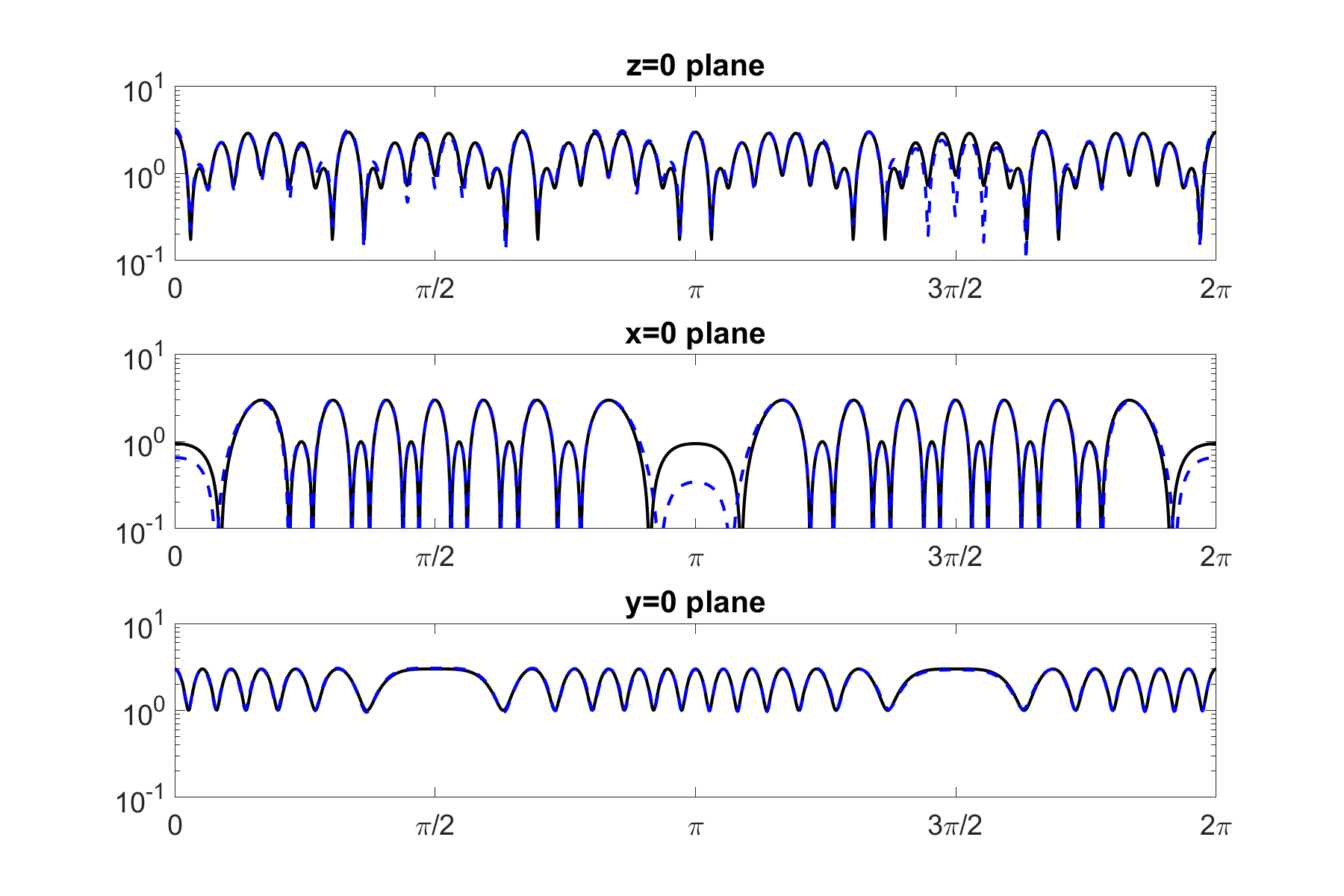}
\caption{Cross-sections of the far-field pattern for the multiple-scattering problem. Black-solid curves are the exact pattern. Blue-dashed curves are the numerical approximation. The wavenumber $k= 2 \pi$.}  
\label{Fig:FFPplanesK2piEx2}
\end{figure}

For this example, we also display some cross-sections of the far-field pattern as shown Figures \ref{Fig:FFPplanesEx2} and \ref{Fig:FFPplanesK2piEx2} for wavenumbers $k=\pi$ and $k=2 \pi$, respectively.

\section{Preconditioner for a boundary element method} 
\label{Section.Precond}

We demonstrate the utility of the proposed local OSRC formulation to precondition the Galerkin boundary element method (BEM) for a multiple scattering problem. This is one of the main applications for OSRCs \cite{Antoine2004,Antoine2005,Antoine2007,Darbas2013,Thierry2014,
Darbas2015,Chaillat2015,Chaillat2017}. The implementation of the Galerkin BEM and the OSRC was done using GYPSILAB, an open source MATLAB library developed by Alouges and 
Aussal \cite{Alouges2018}. We choose a multiple scattering problem in three-dimensional space with two sound-soft unit spheres centered  at $(x,y,z) = (2,1,0)$ and $(x,y,z) = -(2,1,0)$. The incident field $u_{\rm inc}$ is a plane wave of wavenumber $k=4\pi$ propagating along the $x$-axis in the positive direction. We first pose the multiple scattering problem on the single-layer direct integral operator framework \cite{Atkinson-Han-Book-2001,Atkinson1997book,Martin-Book-2006} governed by the following boundary integral equation,
\begin{align} \label{Eqn.BIE_Single}
\begin{bmatrix}
    \SS_{11} & \SS_{12} \\
    \SS_{21} & \SS_{22}
\end{bmatrix} 
\begin{bmatrix}
    \rho_{1} \\
    \rho_{2}
\end{bmatrix}  = 
\begin{bmatrix}
    f_{1} \\
    f_{2}
\end{bmatrix}
\end{align}
where $\rho_{j}$ are the unknown boundary densities on $\partial \Omega_{j}$, $f_{j} = - u_{\rm inc}|_{\partial \Omega_{j}}$ is the Dirichlet data on each sphere $\partial \Omega_{j}$, and the single-layer operators $\SS_{ij}$ are defined as 
\begin{align} \label{Eqn.SingleLayer}
(\SS_{ij} \rho)(\xx) = \int_{\partial \Omega_{j}}  \Phi(\xx,\yy) \rho(\yy) dS(\yy), \qquad \xx \in \partial \Omega_{i}.
\end{align}
At the continuous level, the system (\ref{Eqn.BIE_Single}) can be solved using a block-Jacobi fixed-point iterative method \cite{Atkinson-Han-Book-2001,Atkinson1997book,Thierry2014} of this form
\begin{align} \label{Eqn.BIE_fixed-point}
\begin{bmatrix}
    v_{1} \\
    v_{2}
\end{bmatrix}^{(l+1)} = \begin{bmatrix}
    f_{1} \\
    f_{2}
\end{bmatrix} - \begin{bmatrix}
    \textbf{0}_{11} & \SS_{12} \SS_{22}^{-1} \\
    \SS_{21} \SS_{11}^{-1} & \textbf{0}_{22}
\end{bmatrix} \begin{bmatrix}
    v_{1} \\
    v_{2}
\end{bmatrix}^{(l)}  \quad l=0,1,2, ... L
\end{align} 
where $v_{j} = \SS_{jj} \rho_{j}$. 

We precondition this iterative method using the proposed OSRC formulation (\ref{Eqn.SystemMatrix}) in order to reduced the spectral radius (contraction mapping constant) of the iteration matrix for the fixed-point algorithm. The preconditioned algorithm is written as follows
\begin{align} \label{Eqn.BIE_fixed-point_precond}
\begin{bmatrix}
    w_{1} \\
    w_{2}
\end{bmatrix}^{(l+1)} = \begin{bmatrix}
    f_{1} \\
    f_{2}
\end{bmatrix} + \left( \begin{bmatrix}
    \Id_{11} & \textbf{0}_{12} \\
    \textbf{0}_{21} & \Id_{22}
\end{bmatrix} - \begin{bmatrix}
    \textbf{0}_{11} & \SS_{12} \SS_{22}^{-1} \\
    \SS_{21} \SS_{11}^{-1} & \textbf{0}_{22}
\end{bmatrix} 
\begin{bmatrix}
    \Id_{11} & \PP_{12} \\
    \PP_{21} & \Id_{22}
\end{bmatrix}^{-1} \right)
\begin{bmatrix}
    w_{1} \\
    w_{2}
\end{bmatrix}^{(l)}  \quad l=0,1,2, ... L.
\end{align}
Notice that this iterative method requires the inversion of the system (\ref{Eqn.SystemMatrix}) which is accomplished running the iterative algorithm (\ref{Eqn.FixedPoint}) nested within each iteration of the algorithm (\ref{Eqn.BIE_fixed-point_precond}).

\begin{figure}[H]
\centering
\includegraphics[width=0.45\textwidth, trim={55 10 25 10}]{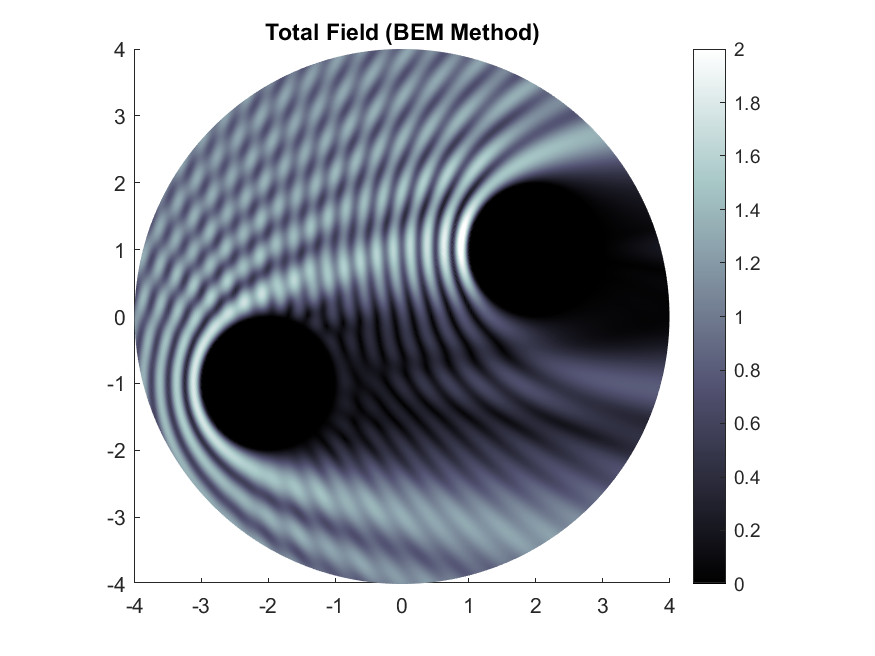}
\includegraphics[width=0.45\textwidth, trim={25 10 55 10}]{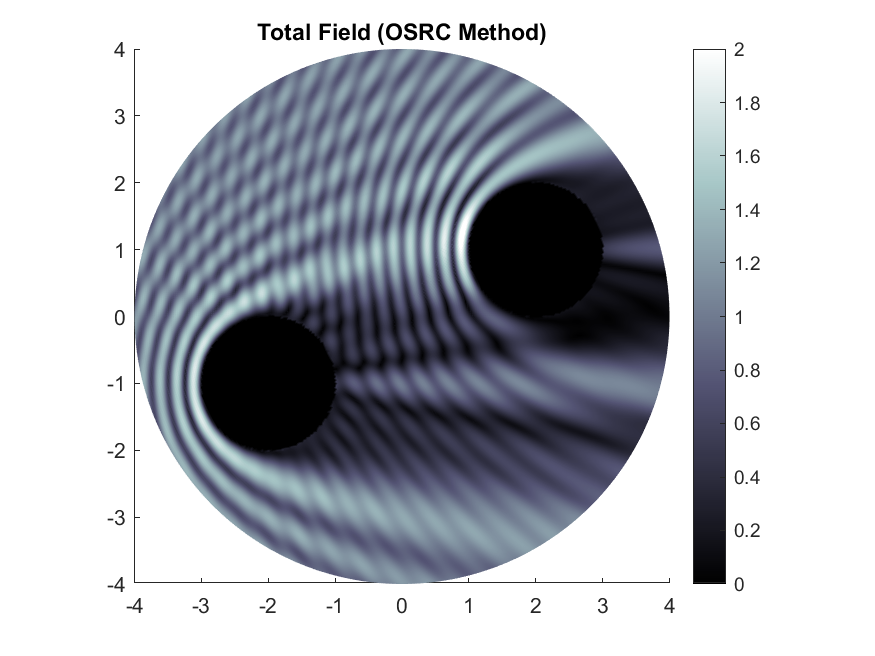}
\caption{Evaluation on the $xy$-plane of the numerical solutions using the BEM and OSRC formulation for the scattering of the plane wave from two sound-soft spheres.}  
\label{Fig:TotalFields}
\end{figure}

\begin{figure}[H]
\centering
\includegraphics[width=0.45\textwidth, trim={55 10 25 10}]{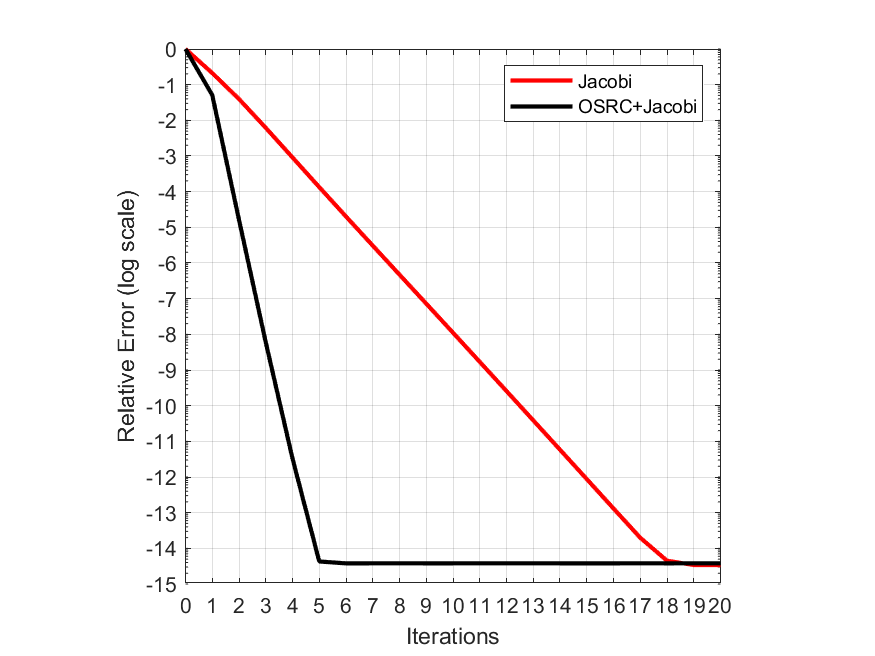}
\includegraphics[width=0.45\textwidth, trim={25 10 55 10}]{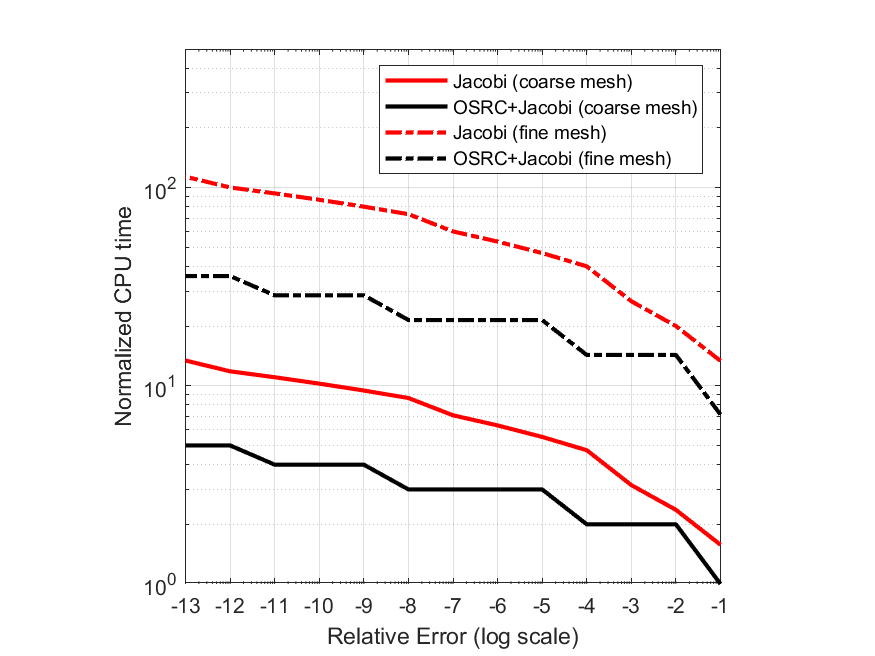}
\caption{Performance and computational cost for the block-Jacobi fixed point iterations (Jacobi) and OSRC-based preconditioned block-Jacobi fixed point iterations (OSRC+Jacobi) to solve the boundary element system. The reported CPU time is normalized to the CPU time needed for one iteration of the preconditioned block-Jacobi method.}  
\label{Fig:Error_BEM}
\end{figure}

For the BEM numerical implementation, we first discretize each sphere using triangular meshes. We ran experiment using two meshes, one with $N=2500$ vertices (coarse mesh) and and the other with $N=5000$ vertices (fine mesh). The operators in (\ref{Eqn.BIE_fixed-point}) and (\ref{Eqn.BIE_fixed-point_precond}) were discretize using the Galerkin projection on continuous piecewise linear functions. The boundary element implementation of the OSRC is easily obtained with the GYPSILAB toolbox. The discretization in the weak form of the propagator (\ref{Eqn.MatrixForm}) corresponds to invert a two-dimensional elliptic system governed by the Laplace-Beltrami operator on each surface, and the function $\AA_{ij}$ in (\ref{Eqn.Entries}) is pre-computed offline on the Gauss quadrature points of the meshes. The numerical approximations for the total acoustic field on the $xy$ cross-sectional plane using the BEM and OSRC formulation are displayed in Figure \ref{Fig:TotalFields}. We note that the result from the OSRC formulation is a good approximation to the BEM solution, but not entirely accurate especially in the shadow regions of the spheres. Hence, instead of using the OSRC formulation to obtain a final solution, we implemented it as a right-preconditioner as shown in the algorithm (\ref{Eqn.BIE_fixed-point_precond}). The convergence pattern and computational cost for both (\ref{Eqn.BIE_fixed-point}) and (\ref{Eqn.BIE_fixed-point_precond}) are displayed in Figure \ref{Fig:Error_BEM}. This procedure reduced the spectral radius of the iteration matrix by a factor of $300$ approximately, which leads to a significant improvement on the rate of convergence due to the OSRC preconditioner. 

For the coarse and fine meshes, the OSRC preconditioner adds about $27\%$ and $7\%$ more computational work, respectively, to each iteration of the block-Jacobi method. This reduction in the relative added computational burden as the mesh is refined is consistent with the linear complexity $\mathcal{O}(N)$ of the OSRC formulation (described in Section \ref{Subsection.Complexity}). This extra marginal cost pays off in terms of faster convergence. See the left panel in Figure \ref{Fig:Error_BEM}. In fact, to reach a desired level of error, the preconditioned method requires less than half of the total computational work needed for the unpreconditioned block-Jacobi method to reach the same error level. See the right panel in Figure \ref{Fig:Error_BEM}.

\section{Final remarks} \label{Section.Conlusion}

We have formulated an OSRC for multiple scattering problems that involves local operations only. Integration over surfaces is avoided which leads to sparse matrices upon discretization and implementation with $\mathcal{O}(N)$ complexity. We expect that this OSRC will provide an inexpensive initial guess or preconditioner to solve multiple scattering problems using boundary integral equations (BIE) or a fast approximate method to explore parameter spaces for optimization problems. The derivation of adequate preconditioners for BIE is a mandatory requirement to solve large scale wave scattering problems because the discretization of BIE leads to notoriously ill-conditioned, non-hermitian dense matrices \cite{Antoine2004,Antoine2005,Antoine2007,Darbas2013,Thierry2014,
Darbas2015,Chaillat2015,Chaillat2017}. In contrast to some algebraic preconditioners for dense systems (such as splitting, incomplete factorization, and sparsification), the OSRC formulation is based on the physics of the problem, including the wave reflections between the multiple obstacles. This renders the OSRC as an effective preconditioner for this specific problem due to its capacity to account for the global interdependence patterns among entries of the system.      

The major limitations of the proposed formulation are similar to those of other OSRCs. This OSRC provides a rough approximation of the solution without a-priori error estimates and lacking a methodical procedure for convergence. Moreover, to obtain closed-form formulas in Section \ref{Section.Propagation}, we adopted several assumptions: (i) convexity of the surfaces bounding the obstacles, (ii) that the principal curvatures of these surfaces are similar, (iii) that their tangential derivatives are small, and (iv) that the frequency is sufficiently high. Therefore, we expect this OSRC to lose accuracy for surfaces with skewed or rapidly changing curvatures, or for low-frequency problems. As a consequence, the OSRC should not be employed as a viable solution procedure, but rather as a fast pre-processing step for preconditioning or exploration.

Possible future work includes the extension to electromagnetic and elastic waves that govern important engineering applications. It is also possible to increase the order of approximation of the Dirichlet--to--Neumann map to improve the accuracy of the OSRC \cite{Acosta2017c}. Here we have considered the second order differential approximation (\ref{Eqn.Lp1}). The major challenge that still remains to be addressed is the formulation of an OSRC and the associated propagation formula for non-convex surfaces. The same is true for piecewise smooth surfaces in order to handle edges and corners. However, recent progress has been reported in this area by Modave, Geuzaine and Antoine \cite{Modave2020}.

\section*{Acknowledgements} \label{Section.Acknowledgements}

This work was partially supported by NSF grant DMS-1712725. The author would like to
thank Texas Children's Hospital for its support and for the research-oriented environment
provided by the Predictive Analytics Laboratory. The author would like to thank the 
anonymous referees for carefully reviewing the manuscript
and for providing helpful recommendations.

\bibliographystyle{elsarticle-num}
\bibliography{Waves}

\end{document}